\documentclass[prd,twocolumn,superscriptaddress,nofootinbib]{revtex4-2}

\usepackage[dvips]{graphicx}
\usepackage{dcolumn}
\usepackage{bm}
\usepackage[dvipsnames]{xcolor}
\usepackage[normalem]{ulem}
\usepackage[errorstop]{feynmp}
\usepackage[compat=1.1.0]{tikz-feynman} 
\usepackage{amsmath,amsfonts,amssymb}
\usepackage[hidelinks]{hyperref}               
\usepackage{cleveref}

\tikzset{graviton/.style={decorate, decoration={snake, amplitude=.4mm, segment length=1.5mm, pre length=.5mm, post length=.5mm}, double}}
\usepackage{slashed}
\usepackage{soul}
\usepackage[utf8]{inputenc} 

\def\d{\mathrm{d}}

\def\A{\mathcal{A}}
\def\L{\mathcal{L}}

\def\O{\mathcal{O}}
\def\vec{\mathbf}
\def\ii{\mathrm{i}}
\def\e{\mathrm{e}}

\def\M{\mathcal{M}}
\def\veck{\vec{k}}
\def\vecp{\vec{p}}
\def\vecq{\vec{q}}
\def\vecx{\vec{x}}

\newcommand{\rom}[1]{\uppercase\expandafter{\romannumeral #1\relax}}

\def\bal#1\eal{\begin{align}#1\end{align}}

\begin{document}

\title{
High-frequency gravitational waves from first-order phase transitions
}

\author{Wen-Yuan Ai}
\email{wenyuan.ai@oeaw.ac.at}

\affiliation{State Key Laboratory of Dark Matter Physics,\\ Tsung-Dao Lee Institute and School of Physics and Astronomy,
Shanghai Jiao Tong University, Shanghai 201210, China}
\affiliation{
Key Laboratory for Particle Astrophysics and Cosmology (MOE) and Shanghai Key Laboratory for Particle Physics and Cosmology, Shanghai Jiao Tong University, Shanghai 201210, China 
}
\affiliation{Marietta Blau Institute for Particle Physics,, Austrian Academy of Sciences,\\ Dominikanerbastei 16, A-1010 Vienna, Austria}

\begin{abstract}
First-order phase transitions in the early Universe are a well-motivated source of gravitational waves (GWs). In this paper, we identify a previously overlooked GW production mechanism: {\it gravitational transition radiation}, arising from graviton emission by particles whose mass changes as they pass through expanding bubble walls. Unlike conventional sources such as bubble collisions or sound waves, this mechanism operates at the microscopic scale set by the Lorentz-contracted wall thickness, leading to GW emission at significantly higher frequencies. The resulting spectrum features a distinctive shape with a peak frequency redshifting to $f_{\rm peak}\sim T_0\sim 10^{10}\,{\rm Hz}$ where $T_0$ is the current temperature of the Universe. This mechanism is generic and is expected to operate similarly for domain walls and other relativistic interfaces.
\end{abstract}

\maketitle

\noindent

\section{Introduction}

The detection of gravitational waves (GWs) by the LIGO/Virgo collaboration in 2015\,\cite{LIGOScientific:2016aoc} has led us to the era of GW astronomy. Future GW observations hold great promise for probing fundamental questions in high-energy physics and for discovering new physics beyond the Standard Model of particle physics and cosmology. It is therefore crucial to theoretically identify all possible sources of GW production. While conventional sources typically involve abrupt, macroscopic classical processes---such as compact object mergers, the collision and coalescence of cosmological defects, or primordial plasma turbulence—GWs can also be generated by microscopic particle processes. These are often described in terms of graviton production. Notable examples include graviton production from the Sun\,\cite{Weinberg:1965nx,Garcia-Cely:2024ujr}, freeze-in production of GWs from the primordial thermal plasma\,\cite{Ghiglieri:2015nfa,Ghiglieri:2020mhm,Ringwald:2020ist,Ghiglieri:2022rfp,Drewes:2023oxg,Montefalcone:2025gxx,Chen:2025try}, gravitational bremsstrahlung during the cosmic reheating\,\cite{Nakayama:2018ptw,Barman:2023ymn,Kanemura:2023pnv,Bernal:2023wus,Tokareva:2023mrt,Hu:2024awd,Xu:2025wjq,Kanemura:2025rct}, leptogenesis\,\cite{Ghoshal:2022kqp,Datta:2024tne,Murayama:2025thw}, or Dark Matter freeze-in production\,\cite{Konar:2025iuk}, and grviton emission from evaporating primordial black holes\,\cite{Dolgov:2000ht,Anantua:2008am,Dolgov:2011cq,Dong:2015yjs,Arbey:2021ysg,Ireland:2023avg,Gross:2024wkl,Choi:2025hqt}. In this article, for the first time, we identify a new mechanism for GW production, arising from a novel phenomenon that we term {\it gravitational transition radiation} (GTR). 

Transition radiation is usually understood as an electromagnetic phenomenon in which a charged particle emits light when crossing an interface that modifies the photon dispersion relation. Radiation can also be produced if the particle itself experiences a sudden change in properties, such as its mass or charge. In cosmology, such conditions naturally arise near phase boundaries, for example, bubble walls or domain walls. The emission is significantly enhanced when the boundary moves relativistically. Analogously, a sudden change in a particle’s mass can induce the emission of gravitons. In this work, we focus on bubble walls from a first-order phase transition (FOPT) as a concrete realisation of this phenomenon and study its associated GW background. However, the mechanism is expected to apply more broadly to other relativistic interfaces, such as domain walls.

On the other hand, GWs generated by FOPTs in the early Universe offer a powerful probe of new physics\,\cite{Hogan:1986dsh,Kosowsky:1992rz,Kosowsky:1992vn,Kamionkowski:1993fg,Caprini:2015zlo,Caprini:2019egz,Athron:2023xlk}. So far, significant attention has been devoted to GWs sourced by bulk hydrodynamic effects such as sound waves\,\cite{Hindmarsh:2013xza,Hindmarsh:2015qta} and turbulence\,\cite{Kamionkowski:1993fg,Kosowsky:2001xp,Caprini:2006jb,Caprini:2009yp}, as well as scalar field bubble collisions\,\cite{Kosowsky:1992rz,Kosowsky:1992vn,Kamionkowski:1993fg,Caprini:2007xq,Huber:2008hg,Inomata:2024rkt}. 
These processes operate at macroscopic scales, with typical GW frequencies tied to the inverse size of the bubbles at collision. In this work, we demonstrate that GTR generates a qualitatively distinct GW signal during the bubble expansion phase of the transition, characterised by a frequency and spectral shape that are dramatically different from those of conventional sources.

\section{Gravitational transition radiation}

\begin{figure}[h]
    \centering
    \includegraphics[width=0.6\linewidth]{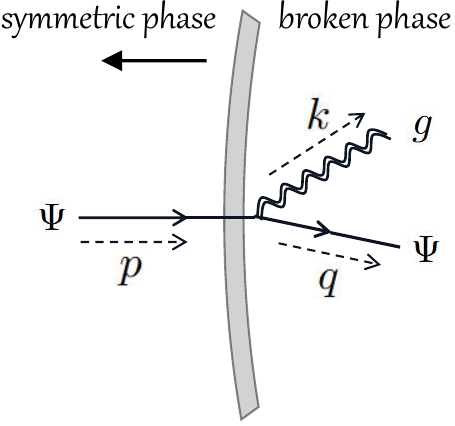}
    \caption{GTR: a particle hitting an expanding wall emits a graviton due to the change in its mass.}
    \label{fig:transition-radiation}
\end{figure}

Consider a scalar field $\Phi$, which may be either fundamental or composite, serving as the order parameter for a FOPT at temperature $T$. This phase transition breaks a certain symmetry and proceeds via the nucleation of bubbles, which separate regions of the broken phase inside the bubble from the surrounding symmetric phase outside. After nucleation, these bubbles will expand until collision. The process of GW production we shall study occurs during the bubble expansion. The bubble field configuration is described by the vacuum expectation value (VEV) of $\Phi$, $\varphi\equiv \langle \Phi\rangle$. Without loss of generality, consider a bubble wall expanding in the negative $z$-direction with a Lorentz factor $\gamma_w\equiv 1/\sqrt{1-v_w^2}$ where $v_w$ is the wall velocity. As the bubble wall expands, particles outside of the bubble impinge on the wall and enter the bubble. A particle $\Psi$ that has a $\varphi$-dependent mass,  which can be a scalar, fermion, or vector boson, can emit a graviton as it passes through the wall due to the change in its mass, leading to the GTR process $\Psi\rightarrow \Psi+ g$, as illustrated in Fig.\,\ref{fig:transition-radiation}.

We write the four-momenta for this process in the rest frame of the wall as
\begin{subequations}
\label{eq:kinematics}
\begin{align}
    &{\rm incoming\
    }\Psi:\quad\  p=(p^0,\vecp_\perp,p^z)\,,\\
    &{\rm outgoing\ }\Psi:\quad\  q=(q^0,\vecq_{\perp},q^z)\,,\\
    &{\rm graviton\
     }  g: \quad\ \ k=(k^0,\veck_{\perp},k^z)\,,
\end{align}
\end{subequations}
where a subscript $\perp$ denotes the direction parallel to the wall and on-shell conditions are understood: $p^0=\sqrt{\vecp_\perp^2+ (p^z)^2 +m^2_\Psi(z)}$ (same for $q$), $k^0=|\veck|$. Since gravitons remain massless in the presence of the wall, one can use $\veck$ to label a free graviton. However, for $\Psi$, the presence of the wall spontaneously breaks the $z$-translation invariance so that $p^0, \vecp_\perp$ (similarly for $q$) are conserved quantities for a one-particle state of $\Psi$ but $p^z$ is $z$-dependent. Although one can use $p^0, \vecp_\perp$ to label one-particle states of $\Psi$, it is more convenient to use $\vecp_s\equiv (\vecp,p^z_s)$ and $\vecp_b\equiv (\vecp,p^z_b)$ to label the incoming and outgoing $\Psi$ particles, respectively. Here, the subscripts ``$s$'' and ``$b$'' denote quantities in the symmetric and broken phases, respectively. For example, $p_{s/b}^z \equiv \sqrt{(p^0)^2-\vecp_\perp^2-m_{\Psi,s/b}^2}$ is the $z$-momentum in the symmetric/broken phase. For simplicity, we take $m_{\Psi, s}=0$ below. The generalisation to the case of $m_{\Psi, s}\neq 0$ is straightforward.

In the wall's rest frame, the flux of $\Psi$ particles impinging on the wall reads
\begin{align}
    J_{\Psi}^{\rm (wall)}= \int\frac{\d^3 \vecp_s}{(2\pi)^3} \frac{p_s^z}{p^0} f^{\rm (wall)}_{\Psi;s}(\vecp_s,T)\,,
\end{align}
where $f^{\rm (wall)}_{\Psi;s}(\vecp_s,T)=1/[\exp(\gamma_w(p^0-v_w p^z)/T)-1]$ is the thermal-equilibrium distribution function of $\Psi$ particles in the symmetric phase in the wall frame.
The graviton energy density in the {\it plasma} frame at the time of generation is given by\,\cite{Ai:2024ikj} (see Appendix\,\ref{sup:BM-method})
\begin{align}
\label{eq:GW-energy-density}
    \rho^{\rm gen}_{\rm GW}&= \frac{1}{v_w\gamma_w}\int\frac{\d^3 \vecp_s}{(2\pi)^3} \frac{p_s^z}{p^0}\times\frac{1}{2 p_s^z}\int\frac{\d^3 \vecq_b}{(2\pi)^3 2 q^0}\int\frac{\d^3 \veck}{(2\pi)^3 2k^0} \notag\\
    &\times \tilde{k}^0\, |\mathcal{M}|^2\, (2\pi)^3\delta(p^0-q^0-k^0)\delta^2(\vecp_\perp-\vecq_\perp-\veck_\perp) \notag\\
    &\times f^{\rm (wall)}_{\Psi;s}(\vecp_s,T)\,[1\pm f^{\rm (wall)}_{\Psi;b}(\vecq_b,T) ]\,,
\end{align}
where $\tilde{k}^0\equiv \gamma_w(k^0-v_w k^z)\approx k^0/2\gamma_w$ is the graviton energy {\it in the plasma frame}, and $\M$ is the invariant transition amplitude for the process $\Psi(p_s)\rightarrow \Psi(q_b)+ g(k)$\,\footnote{Note that here $\M$ is defined from the S-matrix element with a three-dimensional Dirac delta function factored out. As a result, it carries different mass dimensions compared to the conventional definition. For example, in a $1 \to 2$ process, $\M$ is dimensionless, whereas the conventional invariant transition amplitude has mass dimension one.}. The sign in $\pm$ depends on whether $\Psi$ is bosonic or fermionic. Above, we have approximated $1+f_g^{\rm (wall)}(\veck)\approx 1$ on the right-hand side as $f^{\rm (wall)}_g$ is expected to be small at production. (Another argument, as used for gauge boson emission in~\cite{Bodeker:2017cim}, is that the produced graviton can also be absorbed by an incoming $\Psi$ particle. The net effect of including both emission and absorption processes is, approximately, to replace the factor $1+f_g^{\rm (wall)}$ by unity.)

\section{Invariant transition amplitude for GTR}

Because the bubble wall spontaneously breaks the $z$-translation invariance, the transition amplitude $\M$ cannot be computed in the conventional way. One must first construct the mode functions for particles with a $z$-dependent mass. Fortunately, the full procedure has been worked out by Bödeker and Moore~\cite{Bodeker:2017cim}. We provide a comprehensive review of the Bödeker–Moore method in Appendix~\ref{sup:BM-method}. The general form of $\M$ reads 
\begin{align}
    \M= - \left[2\ii p^0 \left(\frac{V_b}{A_b}-\frac{V_s}{A_s}\right)\right]\,,
\end{align}
where $A_{s/b}$ depend on the kinematics and $V_{s/b}$ depend on the specific vertex under consideration. For our case,  
\begin{subequations}
\begin{align}
    &A_s=-\vecp_\perp^2+\frac{x\vecq_\perp^2+(1-x)\veck_\perp^2  }{x(1-x)}\,,\\
    &A_b=-(\vecp_\perp^2+m^2_{\Psi,b})+\frac{x(\vecq_\perp^2+m^2_{\Psi,b})+(1-x)\veck_\perp^2 }{x(1-x)}\,,
\end{align}    
\end{subequations}
where $x=k^0/p^0$.

To get $V_{s/b}$, we consider the minimal coupling between a field $\Psi$ and the graviton $h_{\mu\nu}$,
\begin{align}
\label{eq:vertex-minimal}
    -\L\supset \frac{1}{m_{\rm Pl}}\, h_{\mu\nu} T^{\mu\nu}_{(\Psi)} \equiv \frac{1}{2}\kappa\,  h_{\mu\nu} T_{(\Psi)}^{\mu\nu}\,, 
\end{align}
where $T^{\mu\nu}_{(\Psi)}$ is the energy-momentum tensor of the $\Psi$ field and $m_{\rm Pl}$ is the reduced Planck mass. In Appendix\,\ref{sup:graviton-Feynman-rules}, we review the Feynman rules for the coupling between a graviton and a scalar, fermion, and vector particle. Here, we take a scalar particle $\Psi=\chi$ as an example. Up to the quadratic term in $\chi$  we have
\begin{align*}
    T^{\mu\nu}_{(\chi)}=(\partial^\mu\chi)(\partial^\nu\chi)-\eta^{\mu\nu}\bigg[\frac{1}{2} (\partial_\rho\chi)\partial^\rho\chi -\frac{1}{2}m^2_\chi(z)\chi^2\bigg]\,,
\end{align*} 
where $m^2_\chi(z)=\lambda \varphi^2(z)/2$. The vertex~\eqref{eq:vertex-minimal} then gives the following $V$-function
\begin{align*}
    V(z) = \frac{\kappa}{2} \epsilon^*_{\mu\nu}(k) \left\{p^\mu q^\nu + p^\nu q^\mu -\eta^{\mu\nu}\left[p\cdot q-  m^2_\chi(z)\right]\right\}\,,
\end{align*}
where $\epsilon_{\mu\nu}$ is the graviton polorisation tensor. Then $V_{s/b}\equiv V(z=\pm\infty)$ corresponds to taking $m_\chi(z)$ in $V(z)$ as the mass in the symmetric/broken phase. Substituting the obtained $|\M|^2$ into Eq.\,\eqref{eq:GW-energy-density}, we are prepared to compute the GW energy density at production.

\section{Gravitational wave power spectrum}

Once gravitons are produced at the phase transition temperature $T$, they redshift as radiation. Assuming no entropy generation after the phase transition\,\footnote{This also assumes that there is no exceptionally strong supercooling. Otherwise, the produced GWs would be significantly diluted by the reheating via percolation of bubbles, leading to an additional factor $(g_{\star,s}(T)/g_{\star,s}(T_{\rm RH}))^{4/3} (T/T_{\rm RH})^4\approx (g_{\star,s}(T)/g_{\star,s}(T_{\rm RH}))^{4/3} (1+\alpha_n)^{-1}$, where $T_{\rm RH}$ is the reheating temperature and $\alpha_n$ is the phase transition strength, in the power spectrum. The relation between $k^0$ and the frequency observed today will be modified to $\tilde{k}^0=\left(\frac{g_{\star,s}(T)}{g_{\star,s}(T_0)}\right)^{\frac{1}{3}} \left(\frac{T_{\rm RH}}{T_0}\right)2\pi f$.}, we then have 
\begin{align}
    \rho_{\rm GW}^{\rm today} =\rho_{\rm GW}^{\rm gen} \left(\frac{g_{\star ,s}(T_0)}{g_{\star,s}(T)}\right)^{\frac{4}{3}} \left(\frac{T_0}{T}\right)^4\,,
\end{align}
where $T_0$ is the temperature today, and $g_{\star,s}$ is the number of entropic degrees of freedom. On the other hand, the frequency measured today is related to the graviton energy at production via $\tilde{k}^0 =[a_0/a(T)] 2\pi f $. Substituting the above equations into $\Omega_{\rm GW}=1/\rho_c\,  \d \rho^{\rm today}_{\rm GW}(f)/\d\ln f$ where $\rho_c$ is the critical energy density, we get
\begin{align}
    \Omega_{\rm GW}(f)=&\frac{1}{\rho_c} \left(\frac{g_{\star ,s}(T_0)}{g_{\star,s}(T)}\right)^{\frac{4}{3}} \left(\frac{T_0}{T}\right)^4\notag\\
    &\times\left.\frac{\d\rho_{\rm GW}^{\rm gen}(\tilde{k}^0)}{\d \ln \tilde{k}^0}\right|_{\tilde{k}^0=\left(\frac{g_{\star,s}(T)}{g_{\star,s}(T_0)}\right)^{\frac{1}{3}} \left(\frac{T}{T_0}\right)2\pi f }\,.
\end{align}

\begin{figure}[ht]
    \centering
\includegraphics[width=1\linewidth]{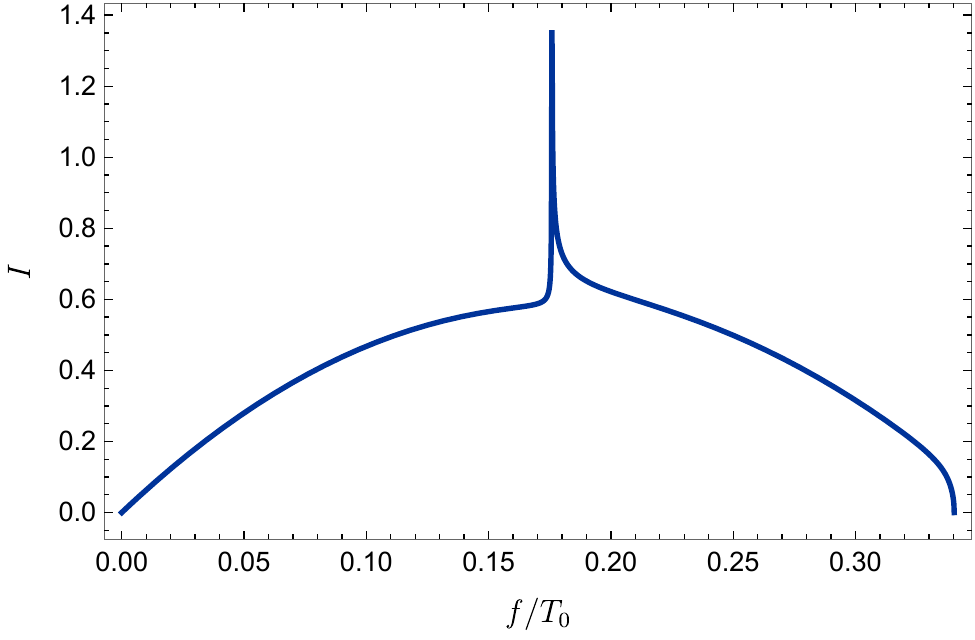}
    \caption{The shape of the power spectrum for $g_{\star,s}(T)=100$. The peak frequency today is at $f_{\rm peak}\approx 0.176 T_0$. }
    \label{fig:I2}
\end{figure}

For simplicity, we still consider GTR from a scalar particle. The generalisation to fermions and vector bosons is straightforward. 
After computing $\d\rho^{\rm gen}_{\rm GW}(\tilde{k}^0)/\d \ln \tilde{k}^0$ (see 
Appendix\,\ref{sec:drhodlnk}), we find, for $\gamma_w T\gg m_{\chi,b}$,
\begin{align}
\label{eq:OmegaGW}
    \Omega_{\rm GW}(f) h^2\simeq  3.6\times 10^{-10} &   \left(\frac{100}{g_{\star,s}(T)}\right)^{\frac{4}{3}}\left(\frac{m_{\chi,b}}{m_{\rm pl}}\right)^2\notag\\
    &\times I\left(\frac{f}{T_0};g_{\star,s}(T)\right)\,.
\end{align}
We note that the factor of $(m_{\chi,b}/m_{\rm pl})^2$ is a universal feature for a single graviton emission or absorption~\,\cite{Weinberg:1965nx,Ai:2025xla}. The shape of the power spectrum is described by the dimensionless integral $I$. This integral actually has additional {\it weak} dependences on $\gamma_w$ and $m_{\chi,b}$. Varying these parameters while keeping $f/T_0$ fixed would slightly alter the shape of $I$, but the results remain of the same order of magnitude. For simplicity, we therefore neglect these weak dependencies. Taking an example, $g_{\star,s}(T)=100$, $m_{\chi,b}=T$ and $\gamma_w=20$, we show it in Fig.\,\ref{fig:I2}. We observe that there are an upper bound for the allowed frequency at $f_{\rm max}\approx 0.34 T_0\approx 1.9\times 10^{10}\,{\rm Hz}$ and a peak at $f_{\rm peak}\approx 0.176 T_0\approx 1\times 10^{10}\,{\rm Hz}$. These features can be understood from the kinematics. First, the energy of the outgoing particle $\chi$ in the broken phase must be larger than its mass, giving $(3\gamma_w T/2-k^0)\geq m_{\chi,b}$ where $3\gamma_w T/2$ is the averaged energy of the incoming $\chi$ particle. This leads to the constraint $k^0\leq k^0_{\rm max} \approx 3\gamma_w T/2$, which gives the maximal frequency. Second, the peak frequency occurs when both the $z$-momentum of the graviton and the outgoing $\chi$ particle can go to zero, such that the integrand in the $I$-integral is resonantly enhanced. This can also be seen from Eq.\,\eqref{eq:GW-energy-density}. In view of that there is the Dirac delta function for the energy conservation, it is convenient to make a change of integral variables: $\d q^z_b/q^0 = \d q^0/ q^z_b$, $\d k^z/k^0= \d k^0/k^z$. Then the integral is resonantly enhanced at $q_b^z=k^z=0$, while when only one of them is zero, we do not observe a local peak showing up. Taking $q_b^z=k^z=0$ in the on-shell conditions, $(k^0)^2=(k^z)^2+|\veck_\perp|^2$ and $(p_s^0-k^0)^2=(q_b^z)^2+|\veck_\perp|^2+m^2_{\chi,b}$ where we have taken $\vecq_\perp= -\veck_\perp$ assuming that in the plasma the average $\vecp_\perp$ is zero, we obtain $k^0=\sqrt{(3\gamma_w T/2-k^0)^2-m^2_{\chi,b}}$. This gives $k_{\rm peak}^0\approx 0.75\gamma_w T$ for $\gamma_w T\gg m_{\chi,b}$. These particular values, $k_{\rm max}^0$ and $k_{\rm peak}^0$, are related to the maximal and peak frequencies via the relation $\frac{k^0}{2\gamma_w}\approx \tilde{k}^0 =\left(\frac{g_{\star,s}(T)}{g_{\star,s}(T_0)}\right)^{\frac{1}{3}} \left(\frac{T}{T_0}\right)2\pi f$.  

It is interesting to note that the GW power spectrum is not explicitly sensitive to $\gamma_w$. However, for most of the $\chi$ particles to enter the bubble, we require $\gamma_w T \gg m_{\chi,b}$. 
For $f\ll f_{\rm peak}$, the $I$-function can be fitted with a linear function. Taking $g_{\star,s}(T)=100$, we find, for $\gamma_w T\gg m_{\chi,b}$,
\begin{align}
    \Omega^{\rm IR}_{\rm GW}(f) h^2\simeq 4\times 10^{-17}\left(\frac{m_{\chi,b}}{m_{\rm pl}}\right)^2 \left(\frac{f}{{\rm kHz}}\right)\,.
\end{align}
Such a linear spectrum is expected to break down in the infrared regime, likely when $\tilde{k}^0(f)\sim H(T)$ with $H$ being the Hubble rate at production.

\begin{figure*}[!t]
    \centering
    \includegraphics[width=0.9\linewidth]{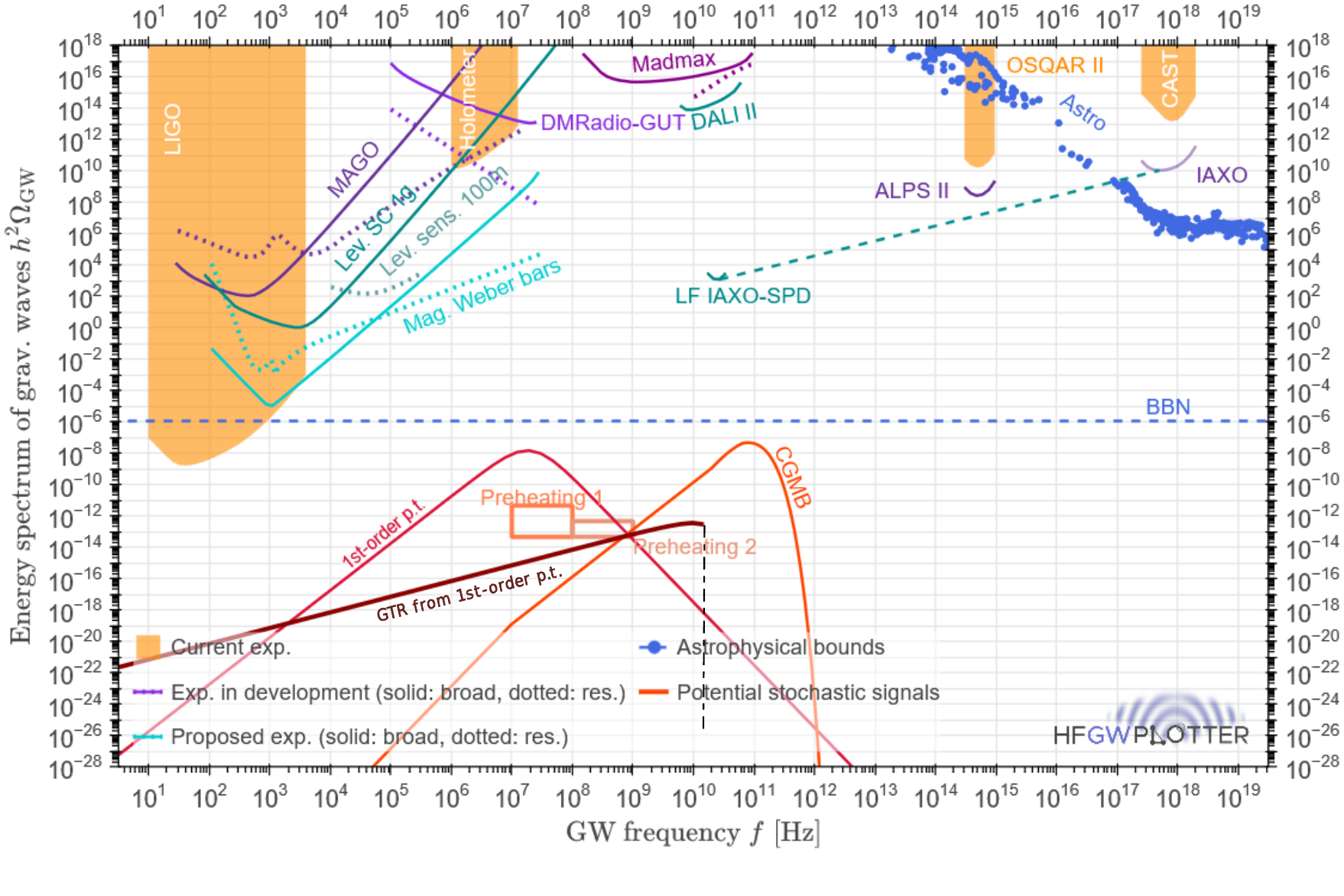}
    \caption{Comparison of the GWs studied in this work (brown) with CGMB (orange) and those from conventional FOPT sources (red), using the benchmark parameters specified in Eq.\,\eqref{eq:BP}. In computing the GW power spectrum, we have multiplied Eq.\,\eqref{eq:OmegaGW} by a factor of 10 to account for multiple degrees of freedom. The figure is generated with the high-frequency GW plotter\,\cite{HFGWPlotter}.}
    \label{fig:example}
\end{figure*}

In Fig.\,\ref{fig:example}, we compare the GTR GWs, represented by the brown curve, with the cosmic gravitational microwave background (CGMB)\,\cite{Ghiglieri:2015nfa,Ghiglieri:2020mhm,Ringwald:2020ist}, represented by the orange curve, and the conventional FOPT GW signals, represented by the red curve, with the following benchmark parameters:
\begin{equation}
\label{eq:BP}
\begin{split}
    &\ T={10^{14}}\,{\rm GeV},\ \gamma_w= {500}, \ m_{\chi,b}={50}\, T,\\ &\ \  T_{\rm max}=m_{\rm pl},\
    \beta/H=20, \ \alpha_n=1,\ v_w\rightarrow 1\,.
\end{split}   
\end{equation}
Above, $T_{\rm max}$ is the maximal temperature ever reached by the Universe, and $\beta/H$ is the dimensionless inverse duration parameter for the phase transition. Also shown are various current experimental bounds (CAST, LIGO, OSQAR II, and Holometer), astrophysical bounds (blue dots), and projected sensitivities of ongoing and proposed experiments. The blue dashed line denotes the BBN bound, while two light-red boxes are envelopes of potential signals from preheating. For more details, see Refs.\,\cite{HFGWPlotter,Aggarwal:2025noe}. Owing to limitations of the interpolated function on a large logarithmic scale, we have manually added a dashed grey line to illustrate the rapid falloff of the GTR GW power spectrum near the maximal frequency. The plot appears to indicate that, like all other cosmological sources of high-frequency GWs, the new signals remain beyond current experimental reach\,\footnote{From the plot, we can see that any high-frequency GWs from cosmological sources that do not violate the BBN bound are beyond the current experimental reach.}. Note, however, that this plot is conservative, and new designs for high-frequency GW detection are actively being developed.

In Fig.~\ref{fig:example}, we see that the new GW signals are relatively suppressed compared with the conventional FOPT GW signals. This may be understood from a quick order-of-magnitude analysis. The peak GW power spectrum from GTR of a scalar particle is given by $\Omega_{\rm GW,peak}^{\rm (GTR)}\sim 10^{-10}(m_{\chi,b}/m_{\rm pl})^2\sim 10^{-13} $. Since we are considering fast bubble walls $v_w\rightarrow 1$, we assume the conventional FOPT GWs are dominated by bubble collisions which gives $\Omega_{\rm GW, peak}^{\rm (bubble)}\sim 10^{-5}(H/\beta)^2\sim 10^{-8} $, where we have used the value of $\beta/H=20$ given in Eq.~\eqref{eq:BP}.

\section{ Discussion}

Observable high-frequency GWs are particularly compelling because they are typically tied to new physics beyond the Standard Model, with perhaps the sole exception being the CGMB if produced minimally from the Standard Model plasma. For a summary of the various known sources, see Ref.\,\cite{Aggarwal:2025noe}. Interest in this area has grown substantially in recent years. There are already interesting proposals and theoretical studies for high-frequency GW detection, including interferometers\,\cite{Ackley:2020atn, Bailes:2019oma, Akutsu:2008qv, Holometer:2016qoh, Nishizawa:2007tn}, microwave and optical cavities\,\cite{Braginskii:1973vm,Grishchuk:1975tg,Braginsky:1979qs,Mensky:2009zz, Caves:1979kq, Pegoraro:1978gv,Pegoraro:1977uv, Reece:1984gv, Reece:1982sc, Ballantini:2005am,Bernard:2002ci,Bernard:2001kp, Ballantini:2003nt, Cruise:2000za, Cruise:2005uq, Cruise:2006zt,Berlin:2021txa,Berlin:2023grv}, optically levitated sensors\,\cite{Aggarwal:2020umq}, mechanical resonators\,\cite{Goryachev:2014nna,Goryachev:2014yra, Aguiar:2010kn, Gottardi:2007zn,Domcke:2024mfu}, superconducting rings\,\cite{Anandan:1982is}, detectors exploiting the inverse Gertsenshtein effect\,\cite{Ejlli:2019bqj,Capdevilla:2024cby,Capdevilla:2025omb}, frequency modulation of photons by GWs\,\cite{Bringmann:2023gba}, and schemes based on the excitation of collective magnon modes~\cite{Ito:2019wcb}. For this reason, it is crucial to identify potential sources of high-frequency GWs.

In this work, we have identified a new source of high-frequency GW production from FOPTs, introducing the GTR mechanism. This mechanism generates a power spectrum with features qualitatively distinct from those of conventional sources. Although detecting such signals---along with other high-frequency GWs---remains challenging in the near future, our work fills an important gap in the understanding of microscopic GW sources and highlights the rich phenomenology associated with relativistic interfaces in the early Universe. In this paper, we have illustrated the mechanism in its simplest realisation, where a scalar particle emits a graviton through minimal gravitational coupling. Exploring how additional structure---such as non-minimal couplings to gravity---could imprint new spectral features is a natural next step. GTR from a vector boson may give a different, potentially larger, GW power spectrum. Another immediate application of our work is to investigate new GW signals sourced by GTR from particles crossing axion domain walls. In this case, the production can last for many Hubble times and hence a larger GW power spectrum is expected. Further, the GTR mechanism may also be applied to astrophysical processes, e.g., a FOPT inside a neutron star\,\cite{Casalderrey-Solana:2022rrn}, where the resulting GWs do not need to respect the BBN upper bound. Our results thus open a new direction for theoretical and phenomenological investigation, and meanwhile, provide a target for future experimental efforts.

{\it Note added}: Shortly after our paper was posted on arXiv, a related study\,\cite{Qiu:2025tmn} appeared, which examines the same process. 

\begin{acknowledgments}
I am grateful to Sebastian A.R. Ellis and Josef Pradler for many useful discussions related to gravitons, and to Carlos Tamarit and Miguel Vanvlasselaer for their helpful comments on this manuscript. I would also like to thank Torsten Bringmann, Gordan Krnjaic, Andrew Long and Maxim Pospelov for comments on this manuscript at the workshop ``From the Cosmos to the Lab: Novel Links and Strategies'', supported by the Mainz Institute for Theoretical Physics (MITP) of the Cluster of Excellence PRISMA$^+$ (Project ID 390831469). This work is supported by the European Union (ERC, NLO-DM, 101044443). 
\end{acknowledgments}

\appendix

\section{B\"{o}deker-Moore method}
\label{sup:BM-method}

In this section, we review the B\"{o}deker-Moore method\,\cite{Bodeker:2017cim} for computing the transition amplitudes in the presence of a planar wall. This method has been used in, e.g., Refs.\,\cite{Hoche:2020ysm,Azatov:2020ufh,Gouttenoire:2021kjv}.
Let us consider the following general interaction
\begin{align}
    -\L \supset  v(z)\,\chi_1\chi_2\chi_3
\end{align}
that can describe the 1-to-2 process $\chi_1(p)\rightarrow \chi_2(q)\chi_3 (k)$. Here we have followed the same notation in Eqs.\,\eqref{eq:kinematics} for the kinematics, with the correspondence $\chi_1(p)\leftrightarrow \Psi(p)$, $\chi_2(q)\leftrightarrow \Psi(q)$, $\chi_3(k)\leftrightarrow g(k)$. Note that the fields and the factor $v(z)$ may carry additional indices reflecting the inner structure in the fields. The factor of $v(z)$ may or may not depend on $z$, depending on whether the VEV $\varphi$ appears in the vertex. For example, one may have an interaction $\lambda \varphi(z)\phi\chi^2$ where $\phi$ is the fluctuation field of $\Phi$ upon the background $\varphi$ (i.e. $\Phi=\varphi(z)+\phi$), and $\chi$ is another scalar field, then $v(z)=\lambda\varphi(z)$. One can also have a Yukawa interaction $y\bar{\psi}\phi\psi$ ($v(z)=y$) or fermion-vector-boson interaction $g_A\bar{\psi}\gamma^\mu A_\mu \psi$ ($v(z)=g_A \gamma^\mu$) where $\gamma^\mu$ are the Dirac gamma matrices.

Even though the VEV $\varphi$ does not appear in the vertex, i.e. $v(z)$ does not depend on $z$, the $z$-translation invariance can still be broken by the $\varphi$-dependent mass terms. As explained in the main text, we shall use $\vecp_s$ to denote the one-particle states $\chi_1$, and $\vecq_b/\veck_b$ to denote the one-particle states of $\chi_2/\chi_3$. (If $\chi_3$ is to be identified as the graviton, then there is no need to use $\veck_b$ since gravitons do not have a $\varphi$-dependent mass.) Taking a scalar field as an example, the free $\chi_1$ field (with the $\varphi$-dependent mass term taken into account) has the following decomposition 
\begin{align}
\label{eq:chi1-decomp}
    \chi_1^{\rm (free)}(x)=&\int\frac{\d^3\vecp_s}{(2\pi)^3\, 2 p^0}\left(\hat{a}_{\vecp_s}\, \zeta_{\vecp_s}(z)\, \e^{-\ii \left(p^0 t -\vecp_\perp \cdot\vecx_\perp\right)}\right. \notag\\
    &\left.+ \hat{a}_{\vecp_s}^\dagger\, \zeta^*_{\vecp_s}(z)\, \e^{\ii \left(p^0 t -\vecp_\perp \cdot\vecx_\perp\right)}\right)\,,
\end{align}
where again $p^0$ should be understood as the on-shell energy.
The momentum labels are also used to distinguish the particle species to simplify the notation. Therefore, $p^0=\sqrt{\vecp_\perp^2+(p^z)^2 +m_{\chi_1}^2 (z)}=\sqrt{\vecp_\perp^2+(p_s^z)^2 +m_{\chi_1,s}^2}$ where $m_{\chi_1,s}$ is the mass of the $\chi_1$ particles in the symmetric phase. Since we consider a process with $\chi_1$ as an incoming particle, the $\chi_1$ distribution function in the symmetric phase $f_{\chi_1;s}^{\rm (wall)}(\vecp_s,T)$ always appears in the expressions of interesting physical quantities. This makes using $\vecp_s$ to label $\chi_1$ particles convenient. The equation of motion for $\zeta_{\vecp_s}$ is obtained by substituting Eq.\,\eqref{eq:chi1-decomp} into the free equation of motion for $\chi_1$. For example, if $\chi_1$ is a scalar field, then we have
\begin{align}
    \left[-(p^0)^2 +\vecp_\perp^2 -\partial_z^2+m^2_{\chi_1}(z)\right]\zeta_{\vecp_s}(z)=0\,.
\end{align}
We require the normalisation for $\zeta_{\vecp_s}$, 
\begin{align}
    \int\d z\, \zeta^*_{\vecp_s}(z) \zeta_{\vecp'_s}(z)=\delta (p^z_s-p'^z_s)\,.
\end{align}
This gives the commutation relations 
\begin{align}
    &[\hat{a}_{\vecp_s}, \hat{a}_{\vecp'_s}]=0\,,\quad    [\hat{a}^\dagger_{\vecp_s}, \hat{a}^\dagger_{\vecp'_s}]=0\,,\notag\\
    &[\hat{a}_{\vecp_s}, \hat{a}^\dagger_{\vecp'_s}]=2 p^0 (2\pi)^3 \delta^3(\vecp_s-\vecp'_s)\,,
\end{align}
and
\begin{align}
    \langle \vecp_s|\vecp'_s\rangle &= 2 p^0 (2\pi)^3 \delta^3(\vecp_s-\vecp'_s)\notag\\
    &= 2p_s^z (2\pi)^3 \delta(p^0-p'^0)\delta^2(\vecp_\perp-\vecp'_\perp)\,.
\end{align}
For $\chi_{2/3}$ free particle states, we have similar expressions but with $\vecp_s$ replaced by $\vecq_b/\veck_b$. 

To study the process $\chi_1(p)\rightarrow \chi_2(q)\chi_3(k)$, we look into the following matrix element
\begin{align}
\label{eq:A}
    &\A_{\chi_1(p)\rightarrow \chi_2(q)\chi_3(k)}=\langle \vecq_b \veck_b |(-\ii) \int\d t\, H_{\rm int}|\vecp_s\rangle\notag\\
    &=-\ii \langle \vecq_b \veck_b |\int\d^4 x\, v(z)\chi_1(x)\chi_2(x)\chi_3(x)|\vecp_s\rangle\,.
\end{align}
The differential probability is then given by
\begin{align}
\label{eq:dP1}
    &\d \mathbb{P}_{\chi_1\rightarrow \chi_2\chi_3}(\vecp_s)=\frac{1}{\langle \vecp_s|\vecp_s\rangle } \int\frac{\d^3 \vecq_{b}}{(2\pi)^3\, 2q^0}[1\pm f_{\chi_2;b}^{(\rm wall)} (\vecq_b,T)]\notag\\
    &\times\int\frac{\d^3 \veck_{b}}{(2\pi)^3\, 2k^0}[1\pm f_{\chi_3;b}^{(\rm wall)} (\veck_b,T)]\, |\A_{\chi_1(p)\rightarrow\chi_2(q)\chi_3(k)}|^2\,.
\end{align}
Above, we have inserted the normalisation factor $\langle \vecp_s|\vecp_s\rangle$. This is a formal procedure. A more rigorous, but slightly more complicated, way is to construct a wave packet state for the incoming $\chi_1$ particle.
Substituting the decomposition of the free fields into Eqs.\,\eqref{eq:dP1} and\,\eqref{eq:A}, we obtain
\begin{align}
\label{eq:dP}
     &\d \mathbb{P}_{\chi_1\rightarrow \chi_2\chi_3}(\vecp_s)=\frac{1}{2p_s^z}\int\frac{\d^3 \vecq_{b}}{(2\pi)^3\, 2q^0}  [1\pm f_{\chi_2;b}^{(\rm wall)} (\vecq_b,T)] \notag\\
     &\times\int\frac{\d^3 \veck_{b}}{(2\pi)^3\, 2k^0}[1\pm f_{\chi_3;b}^{(\rm wall)} (\veck_b,T)](2\pi)^3 \delta(p^0-q^0-k^0)\notag\\
     &\times\delta^2(\vecp_\perp-\vecq_\perp-\veck_\perp) |\M_{\chi_1(p)\rightarrow\chi_2(q)\chi_3(k)}|^2\,,
\end{align}
where
\begin{align}
\label{eq:M}
    \ii \M_{\chi_1(p)\rightarrow\chi_2(q)\chi_3(k)}=&(-\ii) \int\d z\, V(z; \vecp_s,\vecq_b,\veck_b)\notag\\
    &\times\zeta_{\vecp_s}(z) \zeta^*_{\vecq_b}(z)\zeta^*_{\veck_b}(z)\,.
\end{align}
To be general, we have indicated the possible dependence of $V(z;\vecp_s,\vecq_b,\veck_b)$ on the momenta. Occasionally, we will suppress the momenta variables. $V(z)$ can differ from $v(z)$ by the contraction in the inner structure of the fields, e.g., contraction of polarisations. Comparing Eq.\,\eqref{eq:M} with Eq.\,(3.8) of Ref.\,\cite{Bodeker:2017cim}, one may recognise that our $V(z)$ differs from the $V(z)$ in \cite{Bodeker:2017cim} by a minus sign. Also note that a total minus sign in $\M$ will not affect physical results anyway.

A quantity of interest could be the number density of the produced $\chi_3$ particles, i.e., gravitons in the case of gravitational transition radiation. To derive it, we consider a duration $\Delta t$ in the wall frame. In this frame, the flux of $\chi_1$-particles impinging on the wall is
\begin{align}
    J_{\chi_1}^{\rm (wall)}= \int\frac{\d^3 \vecp}{(2\pi)^3} \frac{p_s^z}{p^0} f^{(\rm wall)}_{\chi_1;s}(\vecp_s,T)\,.
\end{align}
The total number of $\chi_3$-particles produced in this time then reads 
\begin{align}
\label{eq:particle-number}
    N_{\chi_3} =& {\rm wall\ area}\times \Delta t \int\frac{\d^3 \vecp_s}{(2\pi)^3} \frac{p_s^z}{p^0} \d\mathbb{P}_{\chi_1\rightarrow \chi_2\chi_3}(\vecp_s)\notag\\&\qquad\qquad\times f^{(\rm wall)}_{\chi_1;s}(\vecp_s,T)\,.
\end{align}
In the plasma frame, the volume swept by the wall is 
\begin{align}
   \Delta V^{\rm (plasma)} = {\rm wall\ area}\times v_w\times \gamma_w\Delta t\,.
\end{align}
Dividing $N_{\chi_3}$ by $\Delta V^{\rm (plasma)}$ then gives the density of produced $\chi_3$-particles in the {\it plasma frame}, 
\begin{align}
\label{eq:production-rate}
    n_{\chi_3}= \frac{1}{v_w\gamma_w}\int\frac{\d^3 \vecp_s}{(2\pi)^3} \frac{p_s^z}{p^0}\d\mathbb{P}_{\chi_1\rightarrow \chi_2\chi_3}(\vecp_s)\times f^{(\rm wall)}_{\chi_1;s}(\vecp_s,T)\,.
\end{align}
The energy density of $\chi_3$-particles can be obtained by inserting a factor of $\tilde{k}^0$, i.e. the graviton energy {\it in the plasma frame}, inside the integrals in $\d \mathbb{P}_{\chi_1\rightarrow\chi_2\chi_3}(\vecp_s)$, Eq.\,\eqref{eq:dP}. This leads to Eq.\,\eqref{eq:GW-energy-density}. 

For relativistic bubble walls ($\gamma_w\gg 1$), and assuming a thermal distribution of $\chi_1$ particles in the symmetric phase
\begin{align}
    f_{\chi_1;s}^{(\rm wall)}(\vecp_s,T)=\frac{1}{\e^{\frac{\gamma_w(p^0-v_w p^z_s)}{T}}-1}\,,
\end{align}
one has that the distribution sharply peaks at $p^z=\gamma_w |\vecp_\perp|\sim \gamma_w T$\,\cite{Ai:2023suz}. Then we are in the WKB regime
\begin{align}
    p^0, q^0, k^0 \sim \gamma_w T \gg \frac{1}{L_w} \sim T\,,
\end{align}
where $L_w$ is the bubble wall width in the wall frame (the proper wall width).
And we have 
\begin{subequations}
\begin{align}
    &\zeta_{\vecp_s}(z)\simeq \sqrt{\frac{p^z_s}{p^z(z)}}\, \e^{\ii\int_0^z\d z'\, p^z(z')}\,,\\
    &\zeta_{\vecq_b}(z)\simeq \sqrt{\frac{q_b^z}{q^z(z)}}\, \e^{\ii\int_0^z\d z'\, q^z(z')}\ (\text{similarly for\ } \veck_b)\,.
\end{align}
\end{subequations}
The above forms satisfy $\zeta_{\vecp_s}(z)\rightarrow \e^{\ii p_s^z z}$ for $z\rightarrow -\infty$ and $\zeta_{\vecq_b}(z)\rightarrow \e^{\ii q_b^z z}$ for $z\rightarrow \infty$.

Using the expansion
\begin{align}
\label{eq:expansion}
    p^z(z')=\sqrt{(p^0)^2-\vecp_\perp^2-m^2_{\chi_1}(z')}\approx p^0-\frac{1}{2}\frac{\vecp_\perp^2+m^2_{\chi_1}(z')}{p^0}\,, 
\end{align}
and similarly for $q^z(z')$ and $k^z(z')$, we have
\begin{align}
\label{eq:Zeta3}
    \zeta_{\vecp_s}(z) \zeta^*_{\vecq_b}(z) \zeta^*_{\veck_b}(z)\approx \e^{\frac{\ii}{2p^0} \int_0^z\d z'\, A(z')}\,,
\end{align}
where
\begin{align}
    A(z')=&-\left(\vecp^2_\perp +m^2_{\chi_1}(z')\right)+\frac{m^2_{\chi_2}(z')+\vecq_\perp^2}{1-x}\notag\\
    &+\frac{m^2_{\chi_3}(z')+\veck^2_\perp}{x}\,.
\end{align}
Above, we have defined $x=k^0/p^0$, $(1-x)=q^0/p^0$ and have used the condition of energy conservation. When doing the expansion\,\eqref{eq:expansion} for $q^z(z')$ and $k^z(z')$, we have assumed 
\begin{align}
    &(p^0)^2 (1-x)^2 \gg \vecq_{\perp}^2+m^2_{\chi_2} \sim T^2\,, \notag\\ & \  (p^0)^2 x^2 \gg \veck_{\perp}^2+m^2_{\chi_3} \sim T^2\notag\\
   &\Rightarrow \  1-x\gg \frac{T}{p^0} \sim\frac{1}{\gamma_w} \ \& \ x \gg \frac{1}{\gamma_w}\,.
\end{align}
As long as $\gamma_w$ is large enough, the phase space violating the above conditions is negligibly small.

The phase in Eq.\,\eqref{eq:Zeta3} is small for $|z|<L_w$. So we can approximate
\begin{align}
    \int\d z'\, A(z')\approx \Theta(-z) A_s\, z+\Theta(z) A_b\, z\,,
\end{align}
where $A_{s,b}$ is the function $A(z')$ evaluated in the symmetric/broken phase. Similarly, $V(z)\approx \Theta(-z) V_s + \Theta (z) V_b$. Substituting these equations into Eq.\,\eqref{eq:M}, one obtains
\begin{align}
    \ii\M= -\ii \left[2\ii p^0 \left(\frac{V_b}{A_b}-\frac{V_s}{A_s}\right)\right]\,. 
\end{align}
Once we know $\M$, we can substitute it into Eq.\,\eqref{eq:production-rate} or \eqref{eq:GW-energy-density} to compute quantities of interest.\\

\noindent {\bf Gravitational transition radiation} For gravitational transition radiation, we do the following replacement
\begin{align}
    \chi_1(p)\rightarrow \Psi(p),\  \chi_2(q)\rightarrow \Psi(q),\ \chi_3(k)\rightarrow g(k)\,. 
\end{align}
For simplicity, we assume $m_{\Psi,s}=0$. Then we have 
\begin{subequations}
\label{eq:As-Ab}
\begin{align}
    &A_s=\frac{-x(1-x)\vecp_\perp^2+x\vecq_\perp^2+(1-x)\veck_\perp^2  }{x(1-x)}\,,\\
    &A_b=\frac{-x(1-x)(\vecp_\perp^2+m^2_{\Psi,b})+x(\vecq_\perp^2+m^2_{\Psi,b})+(1-x)\veck_\perp^2 }{x(1-x)}\,.
\end{align}    
\end{subequations}
In the next Section, we will discuss the function $V(z)$, and therefore $V_{s/b}$, for gravitational transition radiation.

\section{Minimal coupling between \texorpdfstring{$\Psi$}{TEXT} and graviton and Feynman rules}
\label{sup:graviton-Feynman-rules}

\subsection{Gravitational transition radiation from a scalar particle}
Consider the following Lagrangian for a scalar field
\begin{align}
    \L_s/\sqrt{-g} = \frac{1}{2} g^{\mu\nu} (\partial_\mu\chi) \partial_\nu\chi-\frac{1}{2} m^2_\chi(z)\chi^2 \,,
\end{align}
where $m^2_\chi(z)=\lambda \varphi^2(z)/2$.
Doing the expansion $g_{\mu\nu}=\eta_{\mu\nu}+\kappa h_{\mu\nu}$, where $\kappa=2/M_{\rm Pl}$ with $M_{\rm Pl}$ being the reduced Planck mass,
we have $\L_s = \bar{\L}_s+\delta \L_{gs}+\O(\kappa^2)$ where $\bar{\L}_s=\L_s|_{g_{\mu\nu}=\eta_{\mu\nu}}$ and
\begin{align}
    &-\delta\L_{gs}\notag\\
    &=\frac{\kappa}{2} h_{\mu\nu}\left\{(\partial^\mu\chi)(\partial^\nu\chi)-\eta^{\mu\nu}\left[\frac{1}{2} (\partial_\rho\chi)\partial^\rho\chi -\frac{1}{2}m^2_\chi(z)\chi^2\right]\right\}\,.
\end{align} 
This gives the following $V$-function,
\begin{align}
\label{eq:Vgs}
    &V^{(gs)}(z;p,q,k)\notag\\
    &\approx \frac{\kappa}{2} \epsilon^*_{\mu\nu}(k) \left\{p^\mu q^\nu + p^\nu q^\mu -\eta^{\mu\nu}\left[p\cdot q-  m^2_\chi(z)\right]\right\}\,,
\end{align}
where $\epsilon_{\mu\nu}(k)$ is the graviton polorisation tensor. Above, 
\begin{subequations}
\label{eq:pq-definition}
  \begin{align}
    &p=(p^0,\vecp_\perp,\sqrt{(p^0)^2-\vecp_\perp^2-m^2_{\chi}(z)})\\
    &q=(q^0,\vecq_\perp,\sqrt{(q^0)^2-\vecq_\perp^2-m^2_{\chi}(z)})\,,
\end{align}  
\end{subequations}
and it is understood that $p$ is an incoming four-momentum and $q$ is an outgoing four-momentum. Note that in the transverse-traceless gauge, the term proportional to $\eta^{\mu\nu}$ in Eq.\,\eqref{eq:Vgs} would vanish due to $\epsilon_{\mu\nu}\eta^{\mu\nu}=0$. We have checked that removing that term indeed does not change the numerical results given below. The nonvanishing $|\M|^2$ for GTR is therefore due to $A_s\neq A_b$.

Gravitons remain massless in the presence of the bubble wall. So we can use the standard polarisation sum rule for $\epsilon_{\mu\nu}$\,\cite{vanDam:1970vg},
\begin{align}
    \sum_{\lambda=\pm} \epsilon^{\lambda*}_{\mu\nu}(k) \epsilon^{\lambda}_{\alpha\beta}(k)=\frac{1}{2}\left(\hat{\eta}_{\mu\alpha}\hat{\eta}_{\nu\beta}+\hat{\eta}_{\mu\beta}\hat{\eta}_{\nu\alpha}-\hat{\eta}_{\mu\nu}\hat{\eta}_{\alpha\beta}\right)\,,
\end{align}
where 
\begin{align}
\label{eq:hat_eta}
    \hat{\eta}_{\mu\nu}=\eta_{\mu\nu}-\frac{k_{\mu} \bar{k}_{\nu}+\bar{k}_{\mu} k_{\nu}}{k\cdot\bar{k} }\,,
\end{align}
with $\bar{k}=(k^0,-\veck)$. Note that in the presence of the bubble wall, the gravitational Ward identity is violated and therefore one cannot neglect the second term in Eq.\,\eqref{eq:hat_eta} in summing the polarisations of the graviton.

\subsection{Gravitational transition radiation from a fermion}

To derive the interaction between a graviton and a Dirac fermion, one needs to generalise the conventional Dirac Lagrangian in flat spacetime to curved spacetime. For this purpose, one needs to define the Dirac matrices in curved spacetime. This can be done by introducing the vierbein fields\,\cite{Choi:1994ax}. The vector space at any spacetime point is identical to a Minkowski spacetime. Therefore, at any spacetime point, one can introduce an orthonormal basis $\{e^m(x)\}$ (with $m=1,...,4$) where the index is lowered and raised by the Minkowski metric $\eta_{mn}$ ($\eta^{mn}$). 
The vierbein fields are then defined as
\begin{align}
    e_\mu^m(x) \equiv \frac{\partial e^m(x)}{\partial x^\mu}\,.
\end{align}
Then 
\begin{align}
    g_{\mu\nu}(x)= \eta_{mn} e^m_\mu(x) e^n_\nu(x)\,.
\end{align}
The indices $\mu$, $\nu$ are raised and lowered by the curved spacetime metric $g$. For example,  $e_n^\nu(x)= \eta_{mn} g^{\mu\nu}(x)  e^m_\mu (x)$, which is the inverse of $e^n_\mu(x)$. The Dirac matrices in curved spacetime are then defined as $\gamma^\mu(x)= e^\mu_m(x) \gamma^m$. Furthermore, in curved spacetime one has to introduce the covariant derivative on fermionic fields,
\begin{align}
    \nabla_\mu =\partial_\mu +\ii \omega_\mu\,,
\end{align}
where 
\begin{align}
    &\omega_\mu(x)= \frac{1}{4} \sigma^{mn} \left[e^\nu_m (\partial_\mu e_{n\nu}-\partial_\nu e_{n\mu})\right.\notag\\
    &\left.+\frac{1}{2} e^\rho_m e^\sigma_n (\partial_\sigma e_{l\rho}-\partial_\rho e_{l\sigma})e^l_\mu  -(m\leftrightarrow n)\right]\,,
\end{align}
with $\sigma^{mn}=\ii [\gamma^m,\gamma^n]/2$.

Then the Dirac Lagrangian in curved spacetime is 
\begin{align}
    \L_{\rm Dirac}/\sqrt{-g}=\frac{1}{2}\bar{\psi}\left[\ii \gamma^\mu \overset{\rightarrow}{\nabla}_\mu-\ii  \gamma^\mu \overset{\leftarrow}{\nabla}_\mu - 2m_\psi(z) \right]\psi\,,
\end{align}
where $\bar{\psi}\overset{\leftarrow}{\nabla}_\mu= \partial_\mu \bar{\psi}-\ii \bar{\psi}\omega_\mu$, and we have considered a background-field-dependent Dirac mass term $m_\psi(z)=y\varphi(z)$. Expanding the vierbein around the flat spacetime background gives\,\cite{Choi:1994ax}
\begin{align}
    e_\mu^m =\delta^m_\mu + \frac{\kappa}{2} h^m_\mu +\O(\kappa^2)\,.
\end{align}
With all these prepared, we can expand the Dirac Lagrangian around the Minkowski spacetime and obtain the graviton-fermion interaction
\begin{align}
    -\delta \L_{gf}=&\frac{\kappa}{2} h_{\mu\nu}\left\{ \frac{\ii}{4}\bar{\psi}\left[\gamma^\mu\overset{\rightarrow}{\partial^\nu}-\gamma^\mu\overset{\leftarrow}{\partial^\nu} \right]\psi\right.\notag\\
    &\left.-\frac{1}{4}\eta^{\mu\nu} \bar{\psi}\left[\ii \gamma^\rho \overset{\rightarrow}{\partial}_\rho-\ii  \gamma^\rho \overset{\leftarrow}{\partial}_\rho - 2 m_\psi(z)  \right]\psi\right.\notag\\
    &+ (\mu\leftrightarrow \nu)\bigg\}\,.
\end{align}
It is easy to read off the $V$-function from the above equation,
\begin{align}
\label{eq:Vgf}
    V^{(gf)}(z;p,q,k) &\approx  \frac{\kappa}{8} \epsilon^*_{\mu\nu}(k) \bar{u}^{s'}(q) \Big[(p+q)^\mu\gamma^\nu +(p+q)^\nu\gamma^\mu\notag\\
    &-2\eta^{\mu\nu} \left(\slashed{p}  +\slashed{q} -2 m_\psi(z)\right) \Big] u^s(p)  \,\,.
\end{align}
Again, $p$, $q$ are incoming and outgoing four-momenta, respectively, defined in Eqs.\,\eqref{eq:pq-definition}. For anti-fermions, one needs to replace the spinors $u,\bar{u}$ with $v, \bar{v}$, respectively. As commented above, the term with $\eta^{\mu\nu}$ does not contribute in the transverse-traceless gauge. Also note that the spinors are $z$-dependent and hence one should distinguish the spinors in the symmetric phase from those in the broken phase. For a related discussion, see~\cite{Ai:2026}.

\subsection{Gravitational transition radiation from a vector boson}

We consider 
\begin{align}
    \L_{A}/\sqrt{-g}= -\frac{1}{4} (F_{\alpha\beta} \, F^{\alpha\beta})|_g+\frac{1}{2} m^2_A(z)\, g^{\alpha\beta}A_\alpha A_\beta\,,
\end{align}
where  $F_{\alpha\beta}|_g=\nabla_\alpha A_\beta -\nabla_\beta A_\alpha$ and $m^2_A(z)=g^2_A \varphi^2(z)$. This gives the following graviton-vector-boson interaction,
\begin{align}
    -\delta\L_{gA}=\frac{\kappa}{2} h_{\mu\nu}\{-F^{\mu\alpha}F^{\nu}_{\ \alpha}+ m^2_A(z) A^\mu A^\nu  -\eta^{\mu\nu}\L_A|_{g=\eta}\}\,,
\end{align}
where $F_{\mu\nu}$ without $|_g$ is the field strength in flat spacetime. This gives 
\begin{align}
\label{eq:Vgv}
   V^{(gA)}(z;p,q,k)&\approx  \frac{\kappa}{2}\epsilon^*_\beta(q)\epsilon^*_{\mu\nu}(k)\Big[\eta^{\mu\nu}\eta^{\alpha\beta}(p\cdot q-m_A^2(z))\notag\\
   &-\eta^{\mu\nu}p^\beta q^\alpha +\eta^{\nu\alpha}p^\beta q^\mu-\eta^{\alpha\beta}p^\nu q^\mu\notag\\
& +\eta^{\mu\beta} p^\nu q^\alpha
 -\eta^{\nu\alpha}\eta^{\mu\beta}(p\cdot q-m_A^2(z))\notag\\
& +\eta^{\nu\beta} p^\mu q^\alpha-\eta^{\alpha\beta} p^\mu q^\nu +\eta^{\mu\alpha}p^\beta q^\nu\notag\\
&-\eta^{\nu\beta}\eta^{\mu\alpha}\left(p\cdot q-m^2_A(z)\right) \Big]\epsilon_\alpha(p) \,.
\end{align}

\section{Simplify the integral for the produced gravitational wave energy density}

\label{sec:drhodlnk}

It is difficult to do the integral\,\eqref{eq:GW-energy-density} analytically, and so we will rely on numerics. To this end, we simplify the integral\,\eqref{eq:GW-energy-density} first. From the V-functions, Eqs.\,\eqref{eq:Vgs},\,\eqref{eq:Vgf},\,\eqref{eq:Vgv}, and $A_{s/b}$ in Eqs.\,\eqref{eq:As-Ab}, we can obtain $M(p,q,k)$. Since $p$, $q$, $k$ satisfy the on-shell conditions, we can always express $\M$ in terms of $\vecp_s$, $\vecq_b$ and $\veck$. Below, it is more convenient to write $\rho_{\rm GW}^{\rm gen}$ as a function of $k^0$ instead of $\tilde{k}^0$ (recall $\tilde{k}^0\approx k^0/2\gamma_w$). Taking $[1\pm f^{(\rm wall)}_{\Psi;b}(\vecq_b,T)]\approx 1$ and integrating over $\vecq_b$, we obtain
\begin{align}
    &\frac{\d\rho^{\rm gen}_{\rm GW}(k^0)}{\d\ln k^0}= \frac{1}{8\pi v_w\gamma^2_w}\int\frac{\d^3 \vecp_s}{(2\pi)^3 2p^0}f^{\rm (wall)}_{\Psi;s}(\vecp_s,T)\notag\\
    &\times\int\frac{\d^2 \veck_\perp}{(2\pi)^2 }
    \frac{(k^0)^2}{\sqrt{(k^0)^2-\veck_\perp^2}}  \frac{1}{2\sqrt{H(\vecp_s,\veck)}  }\, |\mathcal{M}(\vecp_s,\veck_\perp,k^0)|^2 \,.
\end{align}
where
\begin{align}
    H(\vecp_s,\veck) =  (p^0-k^0)^2-(\vecp_\perp-\veck_\perp)^2-m^2_{\Psi,b}\,,
\end{align}
$\M(\vecp_s,\veck)$ is obtained by taking $\vecq_\perp=\vecp_\perp-\veck_\perp$ and $q_b^z= \sqrt{H(\vecp_s,\veck)}$ in the original expression. Above, we have used $\d k^z=(k^0/k^z) \d k^0$. In principle, we have two solutions for $k^z$ that satisfy the energy-conservation condition, differing from each other by a sign. Given that the incoming particle has a large positive $z$-momentum, we expect that the scattering amplitude receives a dominant contribution with positive $k^z$. Kinematics requires $p^0>k^0$ and $H(\vecp_s,\veck)>0$.

We are interested in the case $\gamma_w\gg 1$ (and so $v_w\rightarrow 1$), since for $\gamma_w\lesssim 1$ the gravitational waves produced from bubble expansion are expected to be subdominant compared with conventional sources. In this case, 
\begin{align}
    f_{\Psi;s}^{\rm (wall)}(\vecp_s,T) &\approx \e^{-\frac{\gamma_w \left( \sqrt{|\vecp_\perp|^2+(p_s^z)^2} - v_w p_s^z\right)}{T}}\notag\\
    &\approx \e^{-\frac{1}{2T}\left(\frac{\gamma_w\vecp_\perp^2 }{p_s^z}+\frac{p_s^z}{\gamma_w} \right) }\,,
\end{align}
where we have used
\begin{align}
    \sqrt{|\vecp_\perp|^2+(p_s^z)^2}\approx p_s^z+\frac{1}{2} \frac{|\vecp_\perp|^2}{p_s^z}\,,\ v_w\approx 1-\frac{1}{2\gamma_w^2}\,.
\end{align}
$f_{\Psi;s}^{(\rm wall)}$, as a function of $p_s^z$, is strongly peaked at $p_s^z=\gamma_w |\vecp_\perp|$. Following Ref.\,\cite{Ai:2023suz}, we can do the integral over $p_s^z$ using the method of steepest descent. 
Denote
\begin{align}
    h(p_s^z)\equiv  -\frac{1}{2T}\left(\frac{\gamma_w|\vecp_\perp|^2}{p_s^z}+\frac{p_s^z}{\gamma_w}\right)\,.
\end{align}
This function is minimal at $p_s^z=\gamma_w|\vecp_\perp|$. Since it appears in the exponent, we can do the Taylor expansion, obtaining
\begin{align}
    h(p_s^z) \approx -\frac{|\vecp_\perp|}{T} - \frac{1}{2} \frac{1}{\gamma_w^2|\vecp_\perp|T} (p_s^z-\gamma_w|\vecp_\perp|)^2\,.
\end{align}
When we do the integral over $p_s^z$, we can simply do the Gaussian integral and let $p_s^z=\gamma_w|\vecp_\perp|$ (and hence $p^0\approx \gamma_w|\vecp_\perp|$) in all other parts. We finally obtain
\begin{align}
\label{eq:drhoGWdk0}
     &\frac{\d\rho^{\rm gen}_{\rm GW}(k^0)}{\d\ln k^0} \approx   \frac{\sqrt{2\pi T}}{512\pi^5\gamma^2_w}\int \d |\vecp_\perp|\,|\vecp_\perp|^{1/2}\e^{-\frac{|\vecp_\perp|}{T}} \int \d |\veck_\perp|\,|\veck_\perp|\notag\\
    &\times\int_0^{2\pi}\d\theta \frac{(k^0)^2}{\sqrt{(k^0)^2-|\veck_\perp|^2}}  \frac{|\left.\M(\vecp_s,\veck)\right|^2_{p_s^z=\gamma_w|\vecp_\perp|}}{\sqrt{\left.H(\vecp_s,\veck)\right|_{p_s^z=\gamma_w|\vecp_\perp|}}}  \,,
\end{align}
where $\theta={\rm arccos}\left(\frac{\vecp_\perp\cdot \veck_\perp}{|\vecp_\perp| |\veck_\perp|}\right)$.

To simplify the integral further, we notice that the integral over $|\vecp_\perp|$ receives the dominant contribution from small $|\vecp_\perp|/T$ due to the exponential function $\exp(-|\vecp_\perp|/T)$ in the integrand. On the other hand, the factor $|\vecp_\perp|^{1/2}$ and the fact that $|\M|^2_{p_b^z=\gamma_w|\vecp_\perp|}$ is regular at $|\vecp_\perp|=0$ indicate that the integral receives negligible contribution as $|\vecp_\perp|\rightarrow 0$. Combining these two observations, one may effectively replace $|\vecp_\perp|$ inside the $\veck$ integral by the following averaged value
\begin{align}
\langle |\vecp_\perp| \rangle =\frac{\int \d |\vecp_\perp|\, |\vecp_\perp|^{3/2} \e^{-\frac{|\vecp_\perp|}{T}} }{\int \d |\vecp_\perp|\, |\vecp_\perp|^{1/2} \e^{-\frac{|\vecp_\perp|}{T}} }  = \frac{3 T}{2}\,.
\end{align}
Mathematically, this amounts to doing a cumulant expansion and taking the leading-order result\,\cite{Ai:2025bjw}. Under such an average and taking into account energy and transverse momentum conservation, we now have 
\begin{subequations}
\label{eq:kinematics2}
\begin{align}
    & p_s = \left( \frac{3\gamma_w T}{2} ,0,0, \frac{3\gamma_w T}{2}\right)\,,\\
     & p_b = \left( \frac{3\gamma_w T}{2} ,0,0, \sqrt{\frac{9\gamma_w^2 T^2}{4} -m_{\Psi,b}^2}\right)\,,\\
     & k = \left(k^0 ,\veck_{\perp},\sqrt{ (k^0)^2 -|\veck_\perp|^2 }\right)\,,\\
    & q_s = \left(\frac{3\gamma_w T}{2}- k^0 ,-\veck_{\perp},\sqrt{\left(\frac{3\gamma_w T}{2}-k^0\right)^2-|\veck_\perp|^2}\right)\,,\\
    & q_b = \left(\frac{3\gamma_w T}{2}- k^0 ,-\veck_{\perp},\sqrt{\left(\frac{3\gamma_w T}{2}-k^0\right)^2-|\veck_\perp|^2-m^2_{\Psi,b}}\right)\,.
\end{align}
\end{subequations}
Note that, although the averaged $|\vecp_\perp|$ is nonzero, we have taken the averaged $\vecp_\perp$ to be zero due to the isotropy in the plane parallel to the wall. This approximation of taking $\vecp_\perp=0$ in computing $\M$ has been used in literature, see e.g.~Refs.\,\cite{Bodeker:2017cim,Azatov:2020ufh}. 

Now $|\M|^2(k^0,|\veck_\perp|)$ becomes a function of $k^0$ and $|\veck_\perp|$, and we can do the integrals over $|\vecp_\perp|$ and $\theta$ in Eq.\,\eqref{eq:drhoGWdk0} trivially. We obtain 
\begin{align}
\label{eq:drhoGWdk02}
     &\frac{\d\rho^{\rm gen}_{\rm GW}(k^0)}{\d\ln k^0} \approx   \frac{\sqrt{2} T^2  }{512\pi^3}\frac{1}{\gamma^2_w}\notag\\
     &\times\int \d |\veck_\perp|\,|\veck_\perp| \frac{(k^0)^2}{\sqrt{(k^0)^2-|\veck_\perp|^2}}\frac{|\left.\M(k^0,|\veck_\perp|)\right|^2}{\sqrt{G(k^0,|\veck_\perp|)}}  \,,
\end{align}
where
\begin{align}
    G(k^0,|\veck_\perp|) = \left(\frac{3\gamma_w T}{2}-k^0\right)^2-|\veck_\perp|^2-m^2_{\Psi,b} \,.
\end{align}
The kinematic constraint gives
\begin{subequations}
\begin{align}
    & k^0 \leq  \frac{3\gamma_w T}{2}-m_{\Psi,b}\,,\\
    &0 \leq |\veck_\perp|\leq {\rm min}\left\{k^0, \sqrt{\left(\frac{3\gamma_w T}{2}-k^0\right)^2 -m_{\Psi,b}^2}\right\}\,.
\end{align}
\end{subequations}

\begin{figure}
    \centering
    \includegraphics[width=1\linewidth]{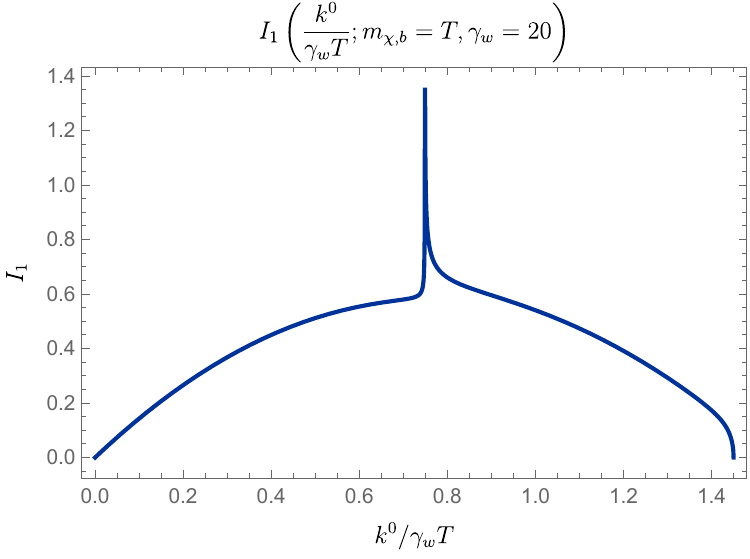}
    \includegraphics[width=1\linewidth]{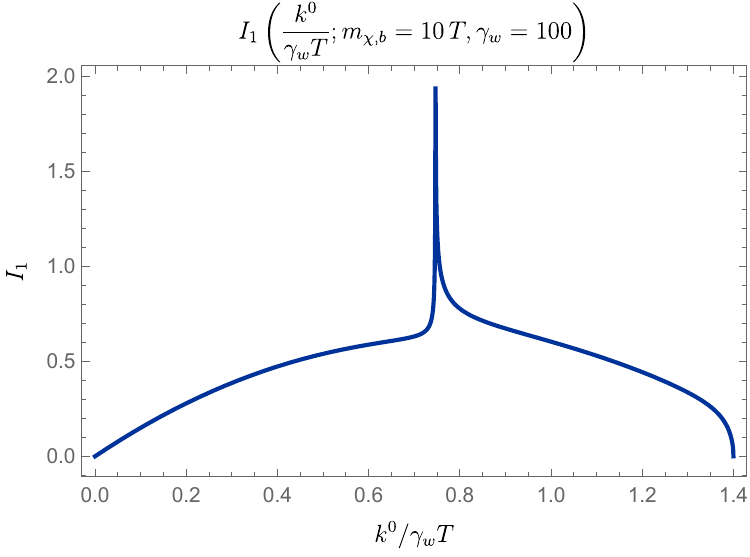}
    \includegraphics[width=1\linewidth]{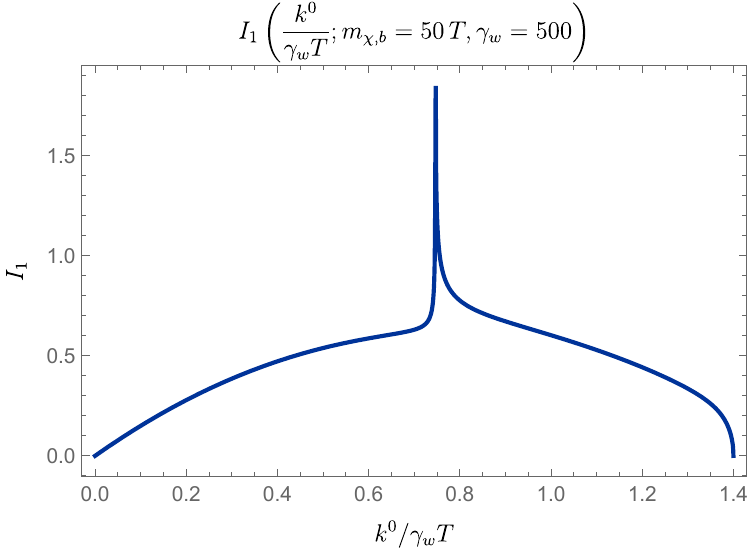}
    \caption{The function $I_1$ in Eq.\,\eqref{eq:drhodlnk0-I} for a few values of $\{m_{\chi,b},\gamma_w\}$. The additional dependence on these two parameters, aside from the dependence on the combination $k^0/\gamma_w T$, is weak. The GW power spectrum has the same shape as $I_1$. }
    \label{fig:I1}
\end{figure}

{\bf Example: Gravitational transition radiation from a scalar particle $\chi$.} Taking
\begin{align}
    x_1=\frac{k^0}{T}\,,\quad x_2=\frac{|\veck_\perp|}{T}\,,\quad x_3=\frac{m_{\chi,b}}{T}\,,
\end{align}
we can write $|\M|^2$ as
\begin{align}
    |\M|^2= \kappa^2 T^2 f(x_1,x_2,x_3,\gamma_w)\,,
\end{align}
where $f$ is a dimensionless function obtained numerically. The final result can be fitted as
\begin{align}
\label{eq:drhodlnk0-I}
    \frac{\d\rho^{\rm gen}_{\rm GW}(k^0)}{\d\ln k^0}\approx & 3.6\times 10^{-4} \left(\frac{m_{\chi,b}}{m_{\rm Pl}}\right)^2 T^4\notag\\
    &\times  I_1\left(\frac{k^0}{\gamma_w T};m_{\chi,b},\gamma_w\right)\,,
\end{align}
where the dependence on $m_{\chi,b}$ and $\gamma_w$ (with $k^0/\gamma_w T$ fixed) in the function $I_1$ is weak. We show and compare the results in Fig.\,\ref{fig:I1} for a few values of $\{m_{\chi,b},\gamma_w\}$.

\bibliographystyle{apsrev4-1}
\bibliography{Ref}

@article{Ai:2025bjw,
    author = "Ai, Wen-Yuan and Carosi, Matthias and Garbrecht, Bjorn and Tamarit, Carlos and Vanvlasselaer, Miguel",
    title = "{Bubble wall dynamics from nonequilibrium quantum field theory}",
    eprint = "2504.13725",
    archivePrefix = "arXiv",
    primaryClass = "hep-ph",
    reportNumber = "MITP-25-029",
    doi = "10.1007/JHEP08(2025)077",
    journal = "JHEP",
    volume = "08",
    pages = "077",
    year = "2025"
}

@unpublished{Ai:2026,
    author = {Ai, Wen-Yuan and Pradler, Josef and Schlachter, Paulina},
    title = "{Fermionic dark matter production from bubble expansion}",
    note = "{\it in preparation}"
}

@article{Ai:2025xla,
    author = "Ai, Wen-Yuan and Ellis, Sebastian A. R. and Pradler, Josef",
    title = "{Soft Gravitons, Hard Truths: Infrared Safety of Particle Processes in a Gravitational-Wave Background}",
    eprint = "2510.27690",
    archivePrefix = "arXiv",
    primaryClass = "hep-ph",
    reportNumber = "UWThPh 2025-19",
    month = "10",
    year = "2025"
}

@article{Qiu:2025tmn,
    author = "Qiu, Dayun and Jiang, Siyu and Huang, Fa Peng",
    title = "{A New Source of Phase Transition Gravitational Waves: Heavy Particle Braking Across Bubble Walls}",
    eprint = "2508.04314",
    archivePrefix = "arXiv",
    primaryClass = "hep-ph",
    month = "8",
    year = "2025"
}

@article{Capdevilla:2025omb,
    author = "Capdevilla, Rodolfo and Harnik, Roni and Kim, Taegyun and Krokotsch, Tom",
    title = "{High-Frequency Gravitational Waves on BREAD}",
    eprint = "2505.21628",
    archivePrefix = "arXiv",
    primaryClass = "hep-ph",
    reportNumber = "FERMILAB-PUB-25-0053-T",
    month = "5",
    year = "2025"
}

@article{Capdevilla:2024cby,
    author = "Capdevilla, Rodolfo and Gelmini, Graciela B. and Hyman, Jonah and Millar, Alexander J. and Vitagliano, Edoardo",
    title = "{Gravitational Wave Detection With Plasma Haloscopes}",
    eprint = "2412.14450",
    archivePrefix = "arXiv",
    primaryClass = "hep-ph",
    reportNumber = "FERMILAB-PUB-24-0436-T",
    month = "12",
    year = "2024"
}

@article{Arbey:2021ysg,
    author = "Arbey, Alexandre and Auffinger, J{\'e}r{\'e}my and Sandick, Pearl and Shams Es Haghi, Barmak and Sinha, Kuver",
    title = "{Precision calculation of dark radiation from spinning primordial black holes and early matter-dominated eras}",
    eprint = "2104.04051",
    archivePrefix = "arXiv",
    primaryClass = "astro-ph.CO",
    reportNumber = "CERN-TH-2021-049",
    doi = "10.1103/PhysRevD.103.123549",
    journal = "Phys. Rev. D",
    volume = "103",
    number = "12",
    pages = "123549",
    year = "2021"
}

@article{Montefalcone:2025gxx,
    author = "Montefalcone, Gabriele and Shams Es Haghi, Barmak and Xu, Tao and Freese, Katherine",
    title = "{Thermal Gravitons from Warm Inflation}",
    eprint = "2507.08739",
    archivePrefix = "arXiv",
    primaryClass = "hep-ph",
    reportNumber = "NORDITA-2025-038, UTWI-19-2025",
    month = "7",
    year = "2025"
}

@article{Murayama:2025thw,
    author = {Murayama, Hitoshi and Noether, Bea and Sch{\"u}tte-Engel, Jan},
    title = "{Observing Leptogenesis in Action with Gravitational Waves}",
    eprint = "2506.15772",
    archivePrefix = "arXiv",
    primaryClass = "hep-ph",
    reportNumber = "RIKEN-iTHEMS-Report-25",
    month = "6",
    year = "2025"
}

@article{Konar:2025iuk,
    author = "Konar, Partha and Show, Sudipta",
    title = "{Unraveling Freeze-in Dark matter through the echoes of gravitational waves}",
    eprint = "2506.08106",
    archivePrefix = "arXiv",
    primaryClass = "hep-ph",
    month = "6",
    year = "2025"
}

@article{Chen:2025try,
    author = "Chen, Quan and Jiang, Siyu and Qiu, Dayun and Chen, Peilin and Huang, Fa Peng",
    title = "{Freeze-in gravitational waves and dark matter in warm inflation}",
    eprint = "2507.13916",
    archivePrefix = "arXiv",
    primaryClass = "hep-ph",
    month = "7",
    year = "2025"
}

@article{Weinberg:1965nx,
    author = "Weinberg, Steven",
    title = "{Infrared photons and gravitons}",
    doi = "10.1103/PhysRev.140.B516",
    journal = "Phys. Rev.",
    volume = "140",
    pages = "B516--B524",
    year = "1965"
}

@article{Garcia-Cely:2024ujr,
    author = {Garc{\'{\i}}a-Cely, Camilo and Ringwald, Andreas},
    title = {Complete Gravitational-Wave Spectrum of the Sun},
    eprint = "2407.18297",
    archivePrefix = "arXiv",
    primaryClass = "hep-ph",
    month = "7",
    year = "2024"
}

@article{Kanemura:2025rct,
    author = "Kanemura, Shinya and Kaneta, Kunio and Nanda, Dibyendu",
    title = "{On the gravitational waves from super massive RHNs produced at preheating}",
    eprint = "2508.00315",
    archivePrefix = "arXiv",
    primaryClass = "hep-ph",
    reportNumber = "OU-HET 1281",
    month = "8",
    year = "2025"
}

@software{HFGWPlotter,
    author = {Francesco Muia and Andreas Ringwald and Carlos Tamarit},
    title = {High frequency gravitational wave plotter: Stochastic signals},
    url = {https://doi.org/10.5281/zenodo.15720443},
    note={Zenodo, DOI: 10.5281/zenodo.15720443}
}

@article{Casalderrey-Solana:2022rrn,
    author = "Casalderrey-Solana, Jorge and Mateos, David and Sanchez-Garitaonandia, Mikel",
    title = "{Mega-Hertz Gravitational Waves from Neutron Star Mergers}",
    eprint = "2210.03171",
    archivePrefix = "arXiv",
    primaryClass = "hep-th",
    month = "10",
    year = "2022"
}

@article{Choi:2025hqt,
    author = "Choi, Ki-Young and Lkhagvadorj, Erdenebulgan and Mahapatra, Satyabrata",
    title = "{Cosmological Origin of the KM3-230213A event and associated Gravitational Waves}",
    eprint = "2503.22465",
    archivePrefix = "arXiv",
    primaryClass = "hep-ph",
    month = "3",
    year = "2025"
}

@article{Gross:2024wkl,
    author = "Gross, Mathieu and Mambrini, Yann and Kpatcha, Essodjolo and Olea-Romacho, Maria Olalla and Roshan, Rishav",
    title = "{Gravitational wave production during reheating: From the inflaton to primordial black holes}",
    eprint = "2411.04189",
    archivePrefix = "arXiv",
    primaryClass = "hep-ph",
    doi = "10.1103/PhysRevD.111.035020",
    journal = "Phys. Rev. D",
    volume = "111",
    number = "3",
    pages = "035020",
    year = "2025"
}

@article{Dolgov:2000ht,
    author = "Dolgov, A. D. and Naselsky, P. D. and Novikov, I. D.",
    title = "{Gravitational waves, baryogenesis, and dark matter from primordial black holes}",
    eprint = "astro-ph/0009407",
    archivePrefix = "arXiv",
    month = "9",
    year = "2000"
}

@article{Ghoshal:2022kqp,
    author = "Ghoshal, Anish and Samanta, Rome and White, Graham",
    title = "{Bremsstrahlung high-frequency gravitational wave signatures of high-scale nonthermal leptogenesis}",
    eprint = "2211.10433",
    archivePrefix = "arXiv",
    primaryClass = "hep-ph",
    doi = "10.1103/PhysRevD.108.035019",
    journal = "Phys. Rev. D",
    volume = "108",
    number = "3",
    pages = "035019",
    year = "2023"
}

@article{Dong:2015yjs,
    author = "Dong, Ruifeng and Kinney, William H and Stojkovic, Dejan",
    title = "{Gravitational wave production by Hawking radiation from rotating primordial black holes}",
    eprint = "1511.05642",
    archivePrefix = "arXiv",
    primaryClass = "astro-ph.CO",
    doi = "10.1088/1475-7516/2016/10/034",
    journal = "JCAP",
    volume = "10",
    pages = "034",
    year = "2016"
}

@article{Dolgov:2011cq,
    author = "Dolgov, Alexander D. and Ejlli, Damian",
    title = "{Relic gravitational waves from light primordial black holes}",
    eprint = "1105.2303",
    archivePrefix = "arXiv",
    primaryClass = "astro-ph.CO",
    doi = "10.1103/PhysRevD.84.024028",
    journal = "Phys. Rev. D",
    volume = "84",
    pages = "024028",
    year = "2011"
}

@article{Bringmann:2023gba,
    author = "Bringmann, Torsten and Domcke, Valerie and Fuchs, Elina and Kopp, Joachim",
    title = "{High-frequency gravitational wave detection via optical frequency modulation}",
    eprint = "2304.10579",
    archivePrefix = "arXiv",
    primaryClass = "hep-ph",
    reportNumber = "CERN-TH-2023-065, MITP-23-017",
    doi = "10.1103/PhysRevD.108.L061303",
    journal = "Phys. Rev. D",
    volume = "108",
    number = "6",
    pages = "L061303",
    year = "2023"
}

@article{Domcke:2024mfu,
    author = "Domcke, Valerie and Ellis, Sebastian A. R. and Rodd, Nicholas L.",
    title = "{Magnets are Weber Bar Gravitational Wave Detectors}",
    eprint = "2408.01483",
    archivePrefix = "arXiv",
    primaryClass = "hep-ph",
    reportNumber = "CERN-TH-2024-132",
    doi = "10.1103/966v-r5fm",
    journal = "Phys. Rev. Lett.",
    volume = "134",
    number = "23",
    pages = "231401",
    year = "2025"
}

@article{Berlin:2023grv,
    author = {Berlin, Asher and Blas, Diego and Tito D'Agnolo, Raffaele and Ellis, Sebastian A. R. and Harnik, Roni and Kahn, Yonatan and Sch{\"u}tte-Engel, Jan and Wentzel, Michael},
    title = "{Electromagnetic cavities as mechanical bars for gravitational waves}",
    eprint = "2303.01518",
    archivePrefix = "arXiv",
    primaryClass = "hep-ph",
    reportNumber = "FERMILAB-PUB-22-892-SQMS-T",
    doi = "10.1103/PhysRevD.108.084058",
    journal = "Phys. Rev. D",
    volume = "108",
    number = "8",
    pages = "084058",
    year = "2023"
}

@article{Berlin:2021txa,
    author = {Berlin, Asher and Blas, Diego and Tito D'Agnolo, Raffaele and Ellis, Sebastian A. R. and Harnik, Roni and Kahn, Yonatan and Sch{\"u}tte-Engel, Jan},
    title = "{Detecting high-frequency gravitational waves with microwave cavities}",
    eprint = "2112.11465",
    archivePrefix = "arXiv",
    primaryClass = "hep-ph",
    reportNumber = "FERMILAB-PUB-21-724-SQMS-T",
    doi = "10.1103/PhysRevD.105.116011",
    journal = "Phys. Rev. D",
    volume = "105",
    number = "11",
    pages = "116011",
    year = "2022"
}

@article{Ballantini:2003nt,
    author = "Ballantini, R. and Bernard, P. and Chiaveri, E. and Chincarini, A. and Gemme, G. and Losito, R. and Parodi, R. and Picasso, E.",
    title = "{A detector of high frequency gravitational waves based on coupled microwave cavities}",
    doi = "10.1088/0264-9381/20/15/316",
    journal = "Class. Quant. Grav.",
    volume = "20",
    pages = "3505--3522",
    year = "2003"
}

@article{Ito:2019wcb,
    author = "Ito, Asuka and Ikeda, Tomonori and Miuchi, Kentaro and Soda, Jiro",
    title = "{Probing GHz gravitational waves with graviton{\textendash}magnon resonance}",
    eprint = "1903.04843",
    archivePrefix = "arXiv",
    primaryClass = "gr-qc",
    reportNumber = "KOBE-COSMO-19-01",
    doi = "10.1140/epjc/s10052-020-7735-y",
    journal = "Eur. Phys. J. C",
    volume = "80",
    number = "3",
    pages = "179",
    year = "2020"
}

@article{Cruise:2000za,
    author = "Cruise, A. M.",
    title = "{An electromagnetic detector for very-high-frequency gravitational waves}",
    doi = "10.1088/0264-9381/17/13/305",
    journal = "Class. Quant. Grav.",
    volume = "17",
    pages = "2525--2530",
    year = "2000"
}

@article{Cruise:2005uq,
    author = "Cruise, A. M. and Ingley, R. M. J.",
    editor = "Jennrich, O.",
    title = "{A correlation detector for very high frequency gravitational waves}",
    doi = "10.1088/0264-9381/22/10/046",
    journal = "Class. Quant. Grav.",
    volume = "22",
    pages = "S479--S481",
    year = "2005"
}

@article{Bernard:2001kp,
    author = "Bernard, P. and Gemme, G. and Parodi, R. and Picasso, E.",
    title = "{A Detector of small harmonic displacements based on two coupled microwave cavities}",
    eprint = "gr-qc/0103006",
    archivePrefix = "arXiv",
    doi = "10.1063/1.1366636",
    journal = "Rev. Sci. Instrum.",
    volume = "72",
    pages = "2428--2437",
    year = "2001"
}

@article{Ballantini:2005am,
    author = "Ballantini, R. and others",
    title = "{Microwave apparatus for gravitational waves observation}",
    eprint = "gr-qc/0502054",
    archivePrefix = "arXiv",
    reportNumber = "INFN-TC-05-05",
    month = "2",
    year = "2005"
}

@article{Bernard:2002ci,
    author = "Bernard, P. and Chincarini, A. and Gemme, G. and Parodi, R. and Picasso, E.",
    title = "{A Detector of gravitational waves based on coupled microwave cavities}",
    eprint = "gr-qc/0203024",
    archivePrefix = "arXiv",
    reportNumber = "INFN-TC-02-03",
    month = "3",
    year = "2002"
}

@article{Reece:1982sc,
    author = "Reece, C. E. and Reiner, P. J. and Melissinos, A. C.",
    editor = "Donaldson, R. and Gustafson, H. Richard and Paige, F. E.",
    title = "{A Detector for High Frequency Gravitational Effects Based on Parametric Conversion at 10 GHz}",
    reportNumber = "UR-832, COO-3065-340, FERMILAB-CONF-82-142",
    journal = "eConf",
    volume = "C8206282",
    pages = "394--402",
    year = "1982"
}

@article{Reece:1984gv,
    author = "Reece, C. E. and Reiner, P. J. and Melissinos, A. C.",
    title = "{OBSERVATION OF 4 X 10**(-17)-CM HARMONIC DISPLACEMENT USING A 10-GHZ SUPERCONDUCTING PARAMETRIC CONVERTER}",
    doi = "10.1016/0375-9601(84)90811-9",
    journal = "Phys. Lett. A",
    volume = "104",
    pages = "341--344",
    year = "1984"
}

@article{Goryachev:2014nna,
    author = "Goryachev, Maxim and Ivanov, Eugene N. and van Kann, Frank and Galliou, Serge and Tobar, Michael E.",
    title = "{Observation of the Fundamental Nyquist Noise Limit in an Ultra-High $Q$-Factor Cryogenic Bulk Acoustic Wave Cavity}",
    eprint = "1410.4293",
    archivePrefix = "arXiv",
    primaryClass = "physics.ins-det",
    doi = "10.1063/1.4898813",
    journal = "Appl. Phys. Lett.",
    volume = "105",
    pages = "153505",
    year = "2014"
}

@article{Goryachev:2014yra,
    author = "Goryachev, Maxim and Tobar, Michael E.",
    title = "{Gravitational Wave Detection with High Frequency Phonon Trapping Acoustic Cavities}",
    eprint = "1410.2334",
    archivePrefix = "arXiv",
    primaryClass = "gr-qc",
    reportNumber = "Erratum: Phys. Rev. D, 108, 129901(E) (2023)",
    doi = "10.1103/PhysRevD.90.102005",
    journal = "Phys. Rev. D",
    volume = "90",
    number = "10",
    pages = "102005",
    year = "2014",
    note = "[Erratum: Phys.Rev.D 108, 129901 (2023)]"
}

@article{Anandan:1982is,
    author = "Anandan, J. and Chiao, R. Y.",
    title = "{GRAVITATIONAL RADIATION ANTENNAS USING THE SAGNAC EFFECT}",
    doi = "10.1007/BF00756213",
    journal = "Gen. Rel. Grav.",
    volume = "14",
    pages = "515--521",
    year = "1982"
}

@article{Aggarwal:2020umq,
    author = "Aggarwal, Nancy and Winstone, George P. and Teo, Mae and Baryakhtar, Masha and Larson, Shane L. and Kalogera, Vicky and Geraci, Andrew A.",
    title = "{Searching for New Physics with a Levitated-Sensor-Based Gravitational-Wave Detector}",
    eprint = "2010.13157",
    archivePrefix = "arXiv",
    primaryClass = "gr-qc",
    doi = "10.1103/PhysRevLett.128.111101",
    journal = "Phys. Rev. Lett.",
    volume = "128",
    number = "11",
    pages = "111101",
    year = "2022"
}

@article{Cruise:2006zt,
    author = "Cruise, A. M. and Ingley, R. M. J.",
    title = "{A prototype gravitational wave detector for 100-MHz}",
    doi = "10.1088/0264-9381/23/22/007",
    journal = "Class. Quant. Grav.",
    volume = "23",
    pages = "6185--6193",
    year = "2006"
}

@article{Ejlli:2019bqj,
    author = "Ejlli, Aldo and Ejlli, Damian and Cruise, Adrian Mike and Pisano, Giampaolo and Grote, Hartmut",
    title = "{Upper limits on the amplitude of ultra-high-frequency gravitational waves from graviton to photon conversion}",
    eprint = "1908.00232",
    archivePrefix = "arXiv",
    primaryClass = "gr-qc",
    doi = "10.1140/epjc/s10052-019-7542-5",
    journal = "Eur. Phys. J. C",
    volume = "79",
    number = "12",
    pages = "1032",
    year = "2019"
}

@article{Gottardi:2007zn,
    author = "Gottardi, L. and de Waard, A. and Usenko, A. and Frossati, G. and Podt, M. and Flokstra, J. and Bassan, M. and Fafone, V. and Minenkov, Y. and Rocchi, A.",
    title = "{Sensitivity of the spherical gravitational wave detector MiniGRAIL operating at 5 K}",
    eprint = "0705.0122",
    archivePrefix = "arXiv",
    primaryClass = "gr-qc",
    doi = "10.1103/PhysRevD.76.102005",
    journal = "Phys. Rev. D",
    volume = "76",
    pages = "102005",
    year = "2007"
}

@article{Aguiar:2010kn,
    author = "Aguiar, Odylio Denys",
    title = "{The Past, Present and Future of the Resonant-Mass Gravitational Wave Detectors}",
    eprint = "1009.1138",
    archivePrefix = "arXiv",
    primaryClass = "astro-ph.IM",
    doi = "10.1088/1674-4527/11/1/001",
    journal = "Res. Astron. Astrophys.",
    volume = "11",
    pages = "1--42",
    year = "2011"
}

@article{Pegoraro:1977uv,
    author = "Pegoraro, F. and Picasso, E. and Radicati, L. A.",
    title = "{On the Operation of a Tunable Electromagnetic Detector for Gravitational Waves}",
    reportNumber = "Print-78-1048 (CERN)",
    doi = "10.1088/0305-4470/11/10/013",
    journal = "J. Phys. A",
    volume = "11",
    pages = "1949--1962",
    year = "1978"
}

@article{Pegoraro:1978gv,
    author = "Pegoraro, F. and Radicati, L. A. and Bernard, P. and Picasso, E.",
    title = "{Electromagnetic Detector for Gravitational Waves}",
    reportNumber = "PRINT-78-0833 (CERN)",
    doi = "10.1016/0375-9601(78)90792-2",
    journal = "Phys. Lett. A",
    volume = "68",
    pages = "165--168",
    year = "1978"
}

@article{Caves:1979kq,
    author = "Caves, C. M.",
    title = "{MICROWAVE CAVITY GRAVITATIONAL RADIATION DETECTORS}",
    doi = "10.1016/0370-2693(79)90227-2",
    journal = "Phys. Lett. B",
    volume = "80",
    pages = "323--326",
    year = "1979"
}

@article{Mensky:2009zz,
    author = "Mensky, M. B. and Rudenko, V. N.",
    title = "{High-frequency gravitational wave detector with electromagnetic-gravitational resonance}",
    doi = "10.1134/S0202289309020133",
    journal = "Grav. Cosmol.",
    volume = "15",
    pages = "167--170",
    year = "2009"
}

@article{Braginsky:1979qs,
    author = "Braginsky, V. B. and Grishchuk, L. P. and Doroshkevich, A. G. and Mensky, M. B. and Novikov, I. D. and Sazhin, M. V. and Zeldovich, Ya. B.",
    title = "{ON THE ELECTROMAGNETIC DETECTION OF GRAVITATIONAL WAVES}",
    doi = "10.1007/BF00759303",
    journal = "Gen. Rel. Grav.",
    volume = "11",
    pages = "407--409",
    year = "1979"
}

@article{Grishchuk:1975tg,
    author = "Grishchuk, L. P. and Sazhin, M. V.",
    title = "{Excitation and Detection of Standing Gravitational Waves}",
    journal = "Pisma Zh. Eksp. Teor. Fiz.",
    volume = "68",
    pages = "1569--1582",
    year = "1975"
}

@article{Braginskii:1973vm,
    author = "Braginskii, V. B. and Grishchuk, L. P. and Doroshkevich, A. G. and Zeldovich, Ya. B. and Novikov, I. D. and Sazhin, M. V.",
    title = "{Electromagnetic detectors of gravitational waves}",
    journal = "Zh. Eksp. Teor. Fiz.",
    volume = "65",
    pages = "1729--1737",
    year = "1973"
}

@article{Nishizawa:2007tn,
    author = "Nishizawa, Atsushi and others",
    title = "{Laser-interferometric Detectors for Gravitational Wave Background at 100 MHz: Detector Design and Sensitivity}",
    eprint = "0710.1944",
    archivePrefix = "arXiv",
    primaryClass = "gr-qc",
    doi = "10.1103/PhysRevD.77.022002",
    journal = "Phys. Rev. D",
    volume = "77",
    pages = "022002",
    year = "2008"
}

@article{Holometer:2016qoh,
    author = "Chou, Aaron S. and others",
    collaboration = "Holometer",
    title = "{MHz Gravitational Wave Constraints with Decameter Michelson Interferometers}",
    eprint = "1611.05560",
    archivePrefix = "arXiv",
    primaryClass = "astro-ph.IM",
    reportNumber = "FERMILAB-PUB-16-449-AE",
    doi = "10.1103/PhysRevD.95.063002",
    journal = "Phys. Rev. D",
    volume = "95",
    number = "6",
    pages = "063002",
    year = "2017"
}

@article{Akutsu:2008qv,
    author = "Akutsu, Tomotada and others",
    title = "{Search for a stochastic background of 100-MHz gravitational waves with laser interferometers}",
    eprint = "0803.4094",
    archivePrefix = "arXiv",
    primaryClass = "gr-qc",
    doi = "10.1103/PhysRevLett.101.101101",
    journal = "Phys. Rev. Lett.",
    volume = "101",
    pages = "101101",
    year = "2008"
}

@article{Bailes:2019oma,
    author = "Bailes, Matthew and others",
    title = "{Ground-Based Gravitational-Wave Astronomy in Australia: 2019 White Paper}",
    eprint = "1912.06305",
    archivePrefix = "arXiv",
    primaryClass = "astro-ph.IM",
    month = "12",
    year = "2019"
}

@article{Ackley:2020atn,
    author = "Ackley, K. and others",
    title = "{Neutron Star Extreme Matter Observatory: A kilohertz-band gravitational-wave detector in the global network}",
    eprint = "2007.03128",
    archivePrefix = "arXiv",
    primaryClass = "astro-ph.HE",
    doi = "10.1017/pasa.2020.39",
    journal = "Publ. Astron. Soc. Austral.",
    volume = "37",
    pages = "e047",
    year = "2020"
}

@article{Ireland:2023avg,
    author = "Ireland, Aurora and Profumo, Stefano and Scharnhorst, Jordan",
    title = "{Primordial gravitational waves from black hole evaporation in standard and nonstandard cosmologies}",
    eprint = "2302.10188",
    archivePrefix = "arXiv",
    primaryClass = "gr-qc",
    doi = "10.1103/PhysRevD.107.104021",
    journal = "Phys. Rev. D",
    volume = "107",
    number = "10",
    pages = "104021",
    year = "2023"
}

@article{Anantua:2008am,
    author = "Anantua, Richard and Easther, Richard and Giblin, John T.",
    title = "{GUT-Scale Primordial Black Holes: Consequences and Constraints}",
    eprint = "0812.0825",
    archivePrefix = "arXiv",
    primaryClass = "astro-ph",
    doi = "10.1103/PhysRevLett.103.111303",
    journal = "Phys. Rev. Lett.",
    volume = "103",
    pages = "111303",
    year = "2009"
}

@article{Athron:2023xlk,
    author = "Athron, Peter and Bal{\'a}zs, Csaba and Fowlie, Andrew and Morris, Lachlan and Wu, Lei",
    title = "{Cosmological phase transitions: From perturbative particle physics to gravitational waves}",
    eprint = "2305.02357",
    archivePrefix = "arXiv",
    primaryClass = "hep-ph",
    doi = "10.1016/j.ppnp.2023.104094",
    journal = "Prog. Part. Nucl. Phys.",
    volume = "135",
    pages = "104094",
    year = "2024"
}

@article{Inomata:2024rkt,
    author = "Inomata, Keisuke and Kamionkowski, Marc and Kasai, Kentaro and Shakya, Bibhushan",
    title = "{Gravitational Waves from Particles Produced from Bubble Collisions in First-Order Phase Transitions}",
    eprint = "2412.17912",
    archivePrefix = "arXiv",
    primaryClass = "astro-ph.CO",
    month = "12",
    year = "2024"
}

@article{Datta:2024tne,
    author = "Datta, Arghyajit and Sil, Arunansu",
    title = "{Probing Leptogenesis through Gravitational Waves}",
    eprint = "2410.01900",
    archivePrefix = "arXiv",
    primaryClass = "hep-ph",
    month = "10",
    year = "2024"
}

@article{Hu:2024awd,
    author = "Hu, Weiyu and Nakayama, Kazunori and Takhistov, Volodymyr and Tang, Yong",
    title = "{Gravitational wave probe of Planck-scale physics after inflation}",
    eprint = "2403.13882",
    archivePrefix = "arXiv",
    primaryClass = "hep-ph",
    reportNumber = "KEK-QUP-2024-0006, TU-1226, KEK-TH-2607, KEK-Cosmo-0341, IPMU24-0007",
    doi = "10.1016/j.physletb.2024.138958",
    journal = "Phys. Lett. B",
    volume = "856",
    pages = "138958",
    year = "2024"
}

@article{Kanemura:2023pnv,
    author = "Kanemura, Shinya and Kaneta, Kunio",
    title = "{Gravitational waves from particle decays during reheating}",
    eprint = "2310.12023",
    archivePrefix = "arXiv",
    primaryClass = "hep-ph",
    reportNumber = "OU--HET--1206, OU-HET-1206",
    doi = "10.1016/j.physletb.2024.138807",
    journal = "Phys. Lett. B",
    volume = "855",
    pages = "138807",
    year = "2024"
}

@article{Drewes:2023oxg,
    author = "Drewes, Marco and Georis, Yannis and Klaric, Juraj and Klose, Philipp",
    title = "{Upper bound on thermal gravitational wave backgrounds from hidden sectors}",
    eprint = "2312.13855",
    archivePrefix = "arXiv",
    primaryClass = "hep-ph",
    doi = "10.1088/1475-7516/2024/06/073",
    journal = "JCAP",
    volume = "06",
    pages = "073",
    year = "2024"
}

@article{LIGOScientific:2016aoc,
    author = "Abbott, B. P. and others",
    collaboration = "LIGO Scientific, Virgo",
    title = "{Observation of Gravitational Waves from a Binary Black Hole Merger}",
    eprint = "1602.03837",
    archivePrefix = "arXiv",
    primaryClass = "gr-qc",
    reportNumber = "LIGO-P150914",
    doi = "10.1103/PhysRevLett.116.061102",
    journal = "Phys. Rev. Lett.",
    volume = "116",
    number = "6",
    pages = "061102",
    year = "2016"
}

@article{Xu:2025wjq,
    author = "Xu, Xun-Jie and Xu, Yong and Yin, Qiqin and Zhu, Junyu",
    title = "{Full-Spectrum Analysis of Gravitational Wave Production from Inflation to Reheating}",
    eprint = "2505.08868",
    archivePrefix = "arXiv",
    primaryClass = "hep-ph",
    reportNumber = "MITP-25-036",
    month = "5",
    year = "2025"
}

@article{Bernal:2023wus,
    author = "Bernal, Nicol{\'a}s and Cl{\'e}ry, Simon and Mambrini, Yann and Xu, Yong",
    title = "{Probing reheating with graviton bremsstrahlung}",
    eprint = "2311.12694",
    archivePrefix = "arXiv",
    primaryClass = "hep-ph",
    reportNumber = "MITP-23-065",
    doi = "10.1088/1475-7516/2024/01/065",
    journal = "JCAP",
    volume = "01",
    pages = "065",
    year = "2024"
}

@article{Tokareva:2023mrt,
    author = "Tokareva, Anna",
    title = "{Gravitational waves from inflaton decay and bremsstrahlung}",
    eprint = "2312.16691",
    archivePrefix = "arXiv",
    primaryClass = "hep-ph",
    doi = "10.1016/j.physletb.2024.138695",
    journal = "Phys. Lett. B",
    volume = "853",
    pages = "138695",
    year = "2024"
}

@article{Nakayama:2018ptw,
    author = "Nakayama, Kazunori and Tang, Yong",
    title = "{Stochastic Gravitational Waves from Particle Origin}",
    eprint = "1810.04975",
    archivePrefix = "arXiv",
    primaryClass = "hep-ph",
    reportNumber = "UT-18-20",
    doi = "10.1016/j.physletb.2018.11.023",
    journal = "Phys. Lett. B",
    volume = "788",
    pages = "341--346",
    year = "2019",
    note = "[Erratum: Phys.Lett.B 839, 137787 (2023)]"
}

@article{Barman:2023ymn,
    author = "Barman, Basabendu and Bernal, Nicol{\'a}s and Xu, Yong and Zapata, {\'O}scar",
    title = "{Gravitational wave from graviton Bremsstrahlung during reheating}",
    eprint = "2301.11345",
    archivePrefix = "arXiv",
    primaryClass = "hep-ph",
    doi = "10.1088/1475-7516/2023/05/019",
    journal = "JCAP",
    volume = "05",
    pages = "019",
    year = "2023"
}

@article{Ringwald:2020ist,
    author = {Ringwald, Andreas and Sch{\"u}tte-Engel, Jan and Tamarit, Carlos},
    title = "{Gravitational Waves as a Big Bang Thermometer}",
    eprint = "2011.04731",
    archivePrefix = "arXiv",
    primaryClass = "hep-ph",
    reportNumber = "DESY 20-187, DESY-20-187, TUM-HEP-1293-20",
    doi = "10.1088/1475-7516/2021/03/054",
    journal = "JCAP",
    volume = "03",
    pages = "054",
    year = "2021"
}

@article{Ghiglieri:2022rfp,
    author = {Ghiglieri, Jacopo and Sch{\"u}tte-Engel, Jan and Speranza, Enrico},
    title = "{Freezing-in gravitational waves}",
    eprint = "2211.16513",
    archivePrefix = "arXiv",
    primaryClass = "hep-ph",
    doi = "10.1103/PhysRevD.109.023538",
    journal = "Phys. Rev. D",
    volume = "109",
    number = "2",
    pages = "023538",
    year = "2024"
}

@article{Ghiglieri:2020mhm,
    author = "Ghiglieri, J. and Jackson, G. and Laine, M. and Zhu, Y.",
    title = "{Gravitational wave background from Standard Model physics: Complete leading order}",
    eprint = "2004.11392",
    archivePrefix = "arXiv",
    primaryClass = "hep-ph",
    doi = "10.1007/JHEP07(2020)092",
    journal = "JHEP",
    volume = "07",
    pages = "092",
    year = "2020"
}

@article{Ghiglieri:2015nfa,
    author = "Ghiglieri, J. and Laine, M.",
    title = "{Gravitational wave background from Standard Model physics: Qualitative features}",
    eprint = "1504.02569",
    archivePrefix = "arXiv",
    primaryClass = "hep-ph",
    doi = "10.1088/1475-7516/2015/07/022",
    journal = "JCAP",
    volume = "07",
    pages = "022",
    year = "2015"
}

@article{Aggarwal:2025noe,
    author = "Aggarwal, Nancy and others",
    title = "{Challenges and Opportunities of Gravitational Wave Searches above 10 kHz}",
    eprint = "2501.11723",
    archivePrefix = "arXiv",
    primaryClass = "gr-qc",
    reportNumber = "CERN-TH-2025-014, DESY-25-007",
    month = "1",
    year = "2025"
}

@article{Ai:2024ikj,
    author = "Ai, Wen-Yuan and Fairbairn, Malcolm and Mimasu, Ken and You, Tevong",
    title = "{Non-thermal production of heavy vector dark matter from relativistic bubble walls}",
    eprint = "2406.20051",
    archivePrefix = "arXiv",
    primaryClass = "hep-ph",
    reportNumber = "KCL-PH-TH/2024-38",
    doi = "10.1007/JHEP05(2025)225",
    journal = "JHEP",
    volume = "05",
    pages = "225",
    year = "2025"
}

@article{Hogan:1986dsh,
    author = "Hogan, C. J.",
    title = "{Gravitational radiation from cosmological phase transitions}",
    doi = "10.1093/mnras/218.4.629",
    journal = "Mon. Not. Roy. Astron. Soc.",
    volume = "218",
    number = "4",
    pages = "629--636",
    year = "1986"
}

@article{vanDam:1970vg,
    author = "van Dam, H. and Veltman, M. J. G.",
    title = "{Massive and massless Yang-Mills and gravitational fields}",
    doi = "10.1016/0550-3213(70)90416-5",
    journal = "Nucl. Phys. B",
    volume = "22",
    pages = "397--411",
    year = "1970"
}

@article{Choi:1994ax,
    author = "Choi, S. Y. and Shim, J. S. and Song, H. S.",
    title = "{Factorization and polarization in linearized gravity}",
    eprint = "hep-th/9411092",
    archivePrefix = "arXiv",
    reportNumber = "KEK-TH-415, HYUPT-94-10, SNUTP-94-03",
    doi = "10.1103/PhysRevD.51.2751",
    journal = "Phys. Rev. D",
    volume = "51",
    pages = "2751--2769",
    year = "1995"
}

@article{Hoche:2020ysm,
    author = {H\"oche, Stefan and Kozaczuk, Jonathan and Long, Andrew J. and Turner, Jessica and Wang, Yikun},
    title = "{Towards an all-orders calculation of the electroweak bubble wall velocity}",
    eprint = "2007.10343",
    archivePrefix = "arXiv",
    primaryClass = "hep-ph",
    reportNumber = "FERMILAB-PUB-20-274-T",
    doi = "10.1088/1475-7516/2021/03/009",
    journal = "JCAP",
    volume = "03",
    pages = "009",
    year = "2021"
}

@article{Azatov:2020ufh,
    author = "Azatov, Aleksandr and Vanvlasselaer, Miguel",
    title = "{Bubble wall velocity: heavy physics effects}",
    eprint = "2010.02590",
    archivePrefix = "arXiv",
    primaryClass = "hep-ph",
    reportNumber = "SISSA 247/2020/FISI",
    doi = "10.1088/1475-7516/2021/01/058",
    journal = "JCAP",
    volume = "01",
    pages = "058",
    year = "2021"
}

@article{Ai:2023suz,
    author = "Ai, Wen-Yuan",
    title = "{Logarithmically divergent friction on ultrarelativistic bubble walls}",
    eprint = "2308.10679",
    archivePrefix = "arXiv",
    primaryClass = "hep-ph",
    reportNumber = "KCL-PH-TH/2023-45",
    doi = "10.1088/1475-7516/2023/10/052",
    journal = "JCAP",
    volume = "10",
    pages = "052",
    year = "2023"
}

@article{Bodeker:2017cim,
    author = "Bodeker, Dietrich and Moore, Guy D.",
    title = "{Electroweak Bubble Wall Speed Limit}",
    eprint = "1703.08215",
    archivePrefix = "arXiv",
    primaryClass = "hep-ph",
    doi = "10.1088/1475-7516/2017/05/025",
    journal = "JCAP",
    volume = "05",
    pages = "025",
    year = "2017"
}

@article{Gouttenoire:2021kjv,
    author = "Gouttenoire, Yann and Jinno, Ryusuke and Sala, Filippo",
    title = "{Friction pressure on relativistic bubble walls}",
    eprint = "2112.07686",
    archivePrefix = "arXiv",
    primaryClass = "hep-ph",
    reportNumber = "DESY-21-147, IFT-UAM/CSIC-21-146",
    doi = "10.1007/JHEP05(2022)004",
    journal = "JHEP",
    volume = "05",
    pages = "004",
    year = "2022"
}

@article{Kosowsky:2001xp,
    author = "Kosowsky, Arthur and Mack, Andrew and Kahniashvili, Tinatin",
    title = "{Gravitational radiation from cosmological turbulence}",
    eprint = "astro-ph/0111483",
    archivePrefix = "arXiv",
    reportNumber = "RAP-334",
    doi = "10.1103/PhysRevD.66.024030",
    journal = "Phys. Rev. D",
    volume = "66",
    pages = "024030",
    year = "2002"
}

@article{Caprini:2009yp,
    author = "Caprini, Chiara and Durrer, Ruth and Servant, Geraldine",
    title = "{The stochastic gravitational wave background from turbulence and magnetic fields generated by a first-order phase transition}",
    eprint = "0909.0622",
    archivePrefix = "arXiv",
    primaryClass = "astro-ph.CO",
    doi = "10.1088/1475-7516/2009/12/024",
    journal = "JCAP",
    volume = "12",
    pages = "024",
    year = "2009"
}

@article{Caprini:2006jb,
    author = "Caprini, Chiara and Durrer, Ruth",
    title = "{Gravitational waves from stochastic relativistic sources: Primordial turbulence and magnetic fields}",
    eprint = "astro-ph/0603476",
    archivePrefix = "arXiv",
    doi = "10.1103/PhysRevD.74.063521",
    journal = "Phys. Rev. D",
    volume = "74",
    pages = "063521",
    year = "2006"
}

@article{Huber:2008hg,
    author = "Huber, Stephan J. and Konstandin, Thomas",
    title = "{Gravitational Wave Production by Collisions: More Bubbles}",
    eprint = "0806.1828",
    archivePrefix = "arXiv",
    primaryClass = "hep-ph",
    doi = "10.1088/1475-7516/2008/09/022",
    journal = "JCAP",
    volume = "09",
    pages = "022",
    year = "2008"
}

@article{Caprini:2007xq,
    author = "Caprini, Chiara and Durrer, Ruth and Servant, Geraldine",
    title = "{Gravitational wave generation from bubble collisions in first-order phase transitions: An analytic approach}",
    eprint = "0711.2593",
    archivePrefix = "arXiv",
    primaryClass = "astro-ph",
    reportNumber = "CERN-PH-TH-2007-206, SACLAY-T07-142",
    doi = "10.1103/PhysRevD.77.124015",
    journal = "Phys. Rev. D",
    volume = "77",
    pages = "124015",
    year = "2008"
}

@article{Kosowsky:1992rz,
    author = "Kosowsky, Arthur and Turner, Michael S. and Watkins, Richard",
    title = "{Gravitational waves from first order cosmological phase transitions}",
    reportNumber = "FERMILAB-PUB-91-333-A-REV, FERMILAB-PUB-91-333-A",
    doi = "10.1103/PhysRevLett.69.2026",
    journal = "Phys. Rev. Lett.",
    volume = "69",
    pages = "2026--2029",
    year = "1992"
}

@article{Hindmarsh:2015qta,
    author = "Hindmarsh, Mark and Huber, Stephan J. and Rummukainen, Kari and Weir, David J.",
    title = "{Numerical simulations of acoustically generated gravitational waves at a first order phase transition}",
    eprint = "1504.03291",
    archivePrefix = "arXiv",
    primaryClass = "astro-ph.CO",
    reportNumber = "HIP-2015-13-TH",
    doi = "10.1103/PhysRevD.92.123009",
    journal = "Phys. Rev. D",
    volume = "92",
    number = "12",
    pages = "123009",
    year = "2015"
}

@article{Hindmarsh:2013xza,
    author = "Hindmarsh, Mark and Huber, Stephan J. and Rummukainen, Kari and Weir, David J.",
    title = "{Gravitational waves from the sound of a first order phase transition}",
    eprint = "1304.2433",
    archivePrefix = "arXiv",
    primaryClass = "hep-ph",
    reportNumber = "HIP-2013-07-TH",
    doi = "10.1103/PhysRevLett.112.041301",
    journal = "Phys. Rev. Lett.",
    volume = "112",
    pages = "041301",
    year = "2014"
}

@article{Caprini:2019egz,
    author = "Caprini, Chiara and others",
    title = "{Detecting gravitational waves from cosmological phase transitions with LISA: an update}",
    eprint = "1910.13125",
    archivePrefix = "arXiv",
    primaryClass = "astro-ph.CO",
    reportNumber = "DESY-19-159, IPPP/19/27, HIP-2019-14/TH, MITP/19-066, IFT-UAM/CSIC-19-139",
    doi = "10.1088/1475-7516/2020/03/024",
    journal = "JCAP",
    volume = "03",
    pages = "024",
    year = "2020"
}

@article{Caprini:2015zlo,
    author = "Caprini, Chiara and others",
    title = "{Science with the space-based interferometer eLISA. II: Gravitational waves from cosmological phase transitions}",
    eprint = "1512.06239",
    archivePrefix = "arXiv",
    primaryClass = "astro-ph.CO",
    reportNumber = "DESY-15-246",
    doi = "10.1088/1475-7516/2016/04/001",
    journal = "JCAP",
    volume = "04",
    pages = "001",
    year = "2016"
}

@article{Kamionkowski:1993fg,
    author = "Kamionkowski, Marc and Kosowsky, Arthur and Turner, Michael S.",
    title = "{Gravitational radiation from first order phase transitions}",
    eprint = "astro-ph/9310044",
    archivePrefix = "arXiv",
    reportNumber = "IASSNS-HEP-93-44, FERMILAB-PUB-93-235-A",
    doi = "10.1103/PhysRevD.49.2837",
    journal = "Phys. Rev. D",
    volume = "49",
    pages = "2837--2851",
    year = "1994"
}

@article{Kosowsky:1992vn,
    author = "Kosowsky, Arthur and Turner, Michael S.",
    title = "{Gravitational radiation from colliding vacuum bubbles: envelope approximation to many bubble collisions}",
    eprint = "astro-ph/9211004",
    archivePrefix = "arXiv",
    reportNumber = "FERMILAB-PUB-92-295-A",
    doi = "10.1103/PhysRevD.47.4372",
    journal = "Phys. Rev. D",
    volume = "47",
    pages = "4372--4391",
    year = "1993"
}

\end{document}